\documentclass[a4paper,12pt]{article}
\usepackage{jcappub} 
\usepackage{cite}
\usepackage{graphicx}
\usepackage{enumerate}
\usepackage{hyperref}
\usepackage{cancel}
\usepackage{color}
\usepackage{graphicx}
\usepackage{amsmath}
\usepackage{amsfonts}
\usepackage{amssymb}
\usepackage{rotating}
\usepackage{booktabs}
\usepackage{xcolor}
\usepackage{soul}

\def\comment#1{}

\title{\boldmath 
Particle-antiparticle oscillation modes crossing horizon:
baryogenesis 
and dark-matter waves
}

\author{She-Sheng Xue}

\affiliation{ICRANet Piazzale della Repubblica, 10 -65122, Pescara, Italy
\\ Physics Department, Sapienza University of Rome, 
Rome, Italy\\INFN, Sezione di Perugia, 
Perugia, Italy
\\ICTP-AP, University of Chinese Academy of Sciences, Beijing, China
}

\emailAdd{xue@icra.it and she-sheng.xue@cern.ch} 

\abstract{
Quantum massive particle and antiparticle pair production and oscillation during reheating result in a holographic and massive pair plasma state. Perturbations in the densities of particles and antiparticles within this plasma form acoustic waves, characterized by symmetric and asymmetric density contrasts. By deriving the acoustic wave equations and identifying the frequencies of the lowest-lying perturbation modes (with zero wave number), the study shows that the wavelengths of these modes, when compared with the horizon size, suggest the possibility of superhorizon crossing during reheating. This crossing leads to particle-antiparticle asymmetry observable by an observer inside the horizon. The decay of massive particles and antiparticles into baryons generates a net baryon number, potentially explaining baryogenesis. 
The calculated baryon number-to-entropy ratio aligns with observational data. 
This crossing also accounts for dark matter particle-antiparticle asymmetry in the present Universe.
The study also explores perturbation modes with nonzero wave numbers, representing dark-matter acoustic waves. These modes exited the horizon and re-entered after recombination, potentially imprinting on the matter power spectrum at large length scales and influencing the formation of large-scale structures and galaxies.
}

\begin{document}
\maketitle
\flushbottom

\section{\bf Introduction}\label{introduction}
In modern cosmology, as described by the $\Lambda$ cold dark matter model ($\Lambda$CDM), several long-standing issues persist, including the cosmological constant $\Lambda$ (dark energy), inflation, reheating, dark matter, and the coincidence and fine-tuning problems. Inflation
\cite{Starobinsky1980, Guth1981, Linde:1981mu, Mukhanov:1982nu, Albrecht1982, Linde1983, Kallosh:2021mnu}
and reheating \cite{Kofman1994, Kofman1997, Shtanov1995, Bassett1998, Tsujikawa1999, Podolsky2002, Allahverdi2010, Amin2012, Amin2014, Adshead2020} are fundamental epochs leading to the hot Big Bang of standard cosmology. 
Massive particle production and decay 
play crucial roles in ending inflation and initiating reheating.
Numerous studies \cite{Parker1973,Starobinsky1982,Ford1987,Kolb1996,
Chung2001,Chung2000,Chung2003,
Chung2005,
Chung2019
} focus on the 
spontaneous gravitational creation of massive particle $X$ and antiparticle $\bar X$ pairs in the early Universe. These particles have masses $M$ larger than the horizon scale $H$. Some of these 
particles are stable and could be candidates for massive dark matter particles, while others are unstable and decay into Standard Model (SM) particles and/or less massive dark-matter particles. 
These massive particles $X$ and antiparticles $\bar X$ are gravitationally created in pairs, persevering the CPT (charge-parity-time) particle-antiparticle symmetry with equal numbers of particles and antiparticles, resulting in 
a net particle number is zero.

The inflation would erase any initial particle-antiparticle asymmetry. 
During reheating, some massive particle-antiparticle pairs annihilate into SM particles, contributing
the leptons and baryons content and increasing entropy. 
After reheating, the Universe evolves adiabatically with conserved entropy. All processes preserve the fundamental CPT symmetry of particles and antiparticles. 
Therefore, the observation of the baryon and anti-baryon asymmetry, i.e., the baryon number-to-entropy ratio 
$n_{_B}/s=  0.864^{+0.016}_{-0.015}\times 10^{-10}$ \cite{Ade2016}, demands an explanation for the baryon asymmetry of the universe, necessitating an understanding of baryogenesis.

In the context of particle physics and cosmology, the absence of initial particle-antiparticle asymmetry leads to the conclusion that the problem of baryon asymmetry requires a dynamic solution characterized by three necessary conditions \cite{Sakharov1967}:
\begin{enumerate}[{(i)}]
\item Baryon number $B$ and lepton number $L$ non-conservation processes are sources for producing baryon and lepton numbers. For instance, in the Grand Unified Theory, particles $X$ decay processes such as $X\rightarrow Y +B(L)$ contribute to this production. Additionally, the violation of the $B+L$ number by the instanton and sphaleron vacuum structure of the SM in particle physics also plays a role;
\item C-symmetry and CP-symmetry violations avoid the $\bar X$ antiparticle processes $\bar X\rightarrow \bar Y +\bar B(\bar L)$ of producing anti-baryon(lepton) number. Namely, if particles $X$ and antiparticles $\bar X$ have the same numbers, they must have different decay rates due to CP-symmetry violations. The complex phase of SM fermion families mixing explicitly breaks the CP-symmetry, as shown in the neutral $K$ and $B$ systems. The CP asymmetry can also be generated by spontaneous symmetry breaking.
\item The system should depart from thermal equilibrium. It occurs when microscopic interaction rates are smaller than the macroscopic evolution rate. Otherwise, the process 
$ X\rightarrow Y + B(\bar L)$ would be compensated by the inverse process 
$Y + B(\bar L)\rightarrow X$ and therefore the total net baryon number remains zero. 
\end{enumerate}
We recall that these Sakharov conditions are satisfied in the standard cosmology model and the SM of particle physics. However, the effect on baryogenesis is insufficient to match the observed baryon number-to-entropy ratio.  
Particle physicists have proposed numerous ideas to explain baryon asymmetry, incorporating elements beyond the SM \cite{Dolgov1982, Abbott1982, Kuzmin1985, Dine2003}. 
One notable approach is the dynamic process of electroweak baryogenesis, which involves
the first-order phase transition of electroweak symmetry breaking of the SM nontrivial vacuum \cite{Trodden_1999, Morrissey_2012}. In addition, several studies have explored intriguing connections between reheating and 
baryogenesis \cite{Dolgov1995, Dolgov1997, GarciaBellido1999, Davidson2000, Megevand2001, Tranberg2003, Tranberg2006,  Hertzberg2014, Hertzberg2014a, Hertzberg2014b, Lozanov2014, Zurek2014}.

In this article, we do not consider an alternative dynamic solution for baryogenesis following the Sakharov conditions by introducing explicit or spontaneous breaking of the CPT symmetry of particles and antiparticles in Lagrangian or ground state in the Universe evolution. 
Instead, we attempt to explain baryogenesis by studying the horizon-crossing dynamics that possibly cause the different number between massive particles $X$ and antiparticles $\bar X$ pairs, 
which are equally produced in the reheating epoch. Such a difference provided an initial value of particle and antiparticle asymmetry that preserves the entire Universe after the reheating, possibly accounting for baryogenesis and dark matter particle and antiparticle asymmetry. We briefly describe the horizon-crossing dynamics below, and detailed discussions and equations will be in the main text.

In conjunction with the rapidly oscillating component $H_{\rm fast}$ of the horizon, the production and oscillation of particle $X$ and antiparticle $\bar X$ pairs
maintains the fundamental CPT symmetry. It does not produce different numbers of particles and antiparticles, i.e.,  
particle and antiparticle asymmetry.  
However, these $X$ particles and $\bar X$ antiparticles undergo pair oscillations with each other, and their local and instantaneous distributions (densities) 
are different (not in phase) in spacetime. 
Moreover, the pair oscillation frequency or 
wavelength depends on the horizon evolution 
in reheating. As a consequence, two possible situations can occur. 

First, if the wavelength of pair oscillating modes is smaller than the horizon size. 
The local difference of particle and antiparticle distributions (densities) would tend to average to zero across different patches inside the horizon. All particles and antiparticles are inside the horizon, and particle and antiparticle asymmetry does not occur.
Second, if pair oscillating modes cross out the horizon when their wavelength exceeds the size of the horizon. In other words, such super-horizon crossing occurs when the horizon expansion rate is smaller than the oscillating frequency of particle and antiparticle modes.  
Such super-horizon modes freeze outside the horizon and never return to the horizon. Consequently, 
the local difference in particle and antiparticle distributions (densities) would not 
tend to average to zero across different patches inside the horizon. The different numbers of particles and antiparticles appear inside the horizon, resulting in a net $X$ and $\bar X$ particle number, 
namely the particle and antiparticle asymmetry for observers inside the horizon. In both situations, the total sum of $X$ and $\bar X$ particle numbers inside and outside the horizon remains zero, as required by particle number conservation during their production and oscillation.

Moreover, through massive particles (pairs) decay (annihilate) to SM particles within the horizon, such 
an $X-\bar X$ asymmetry in the reheating could be an initial source for the asymmetry between SM particles and antiparticles, namely the lepton and baryon numbers $B+L$ violation while conserving $B-L$ and electric charge in the present Universe.

We will investigate the horizon-crossing dynamics and phenomena potentially responsible for producing the initial particle and antiparticle asymmetry inside the horizon during the reheating epoch and how this relates to the observed baryogenesis. Notably, the horizon-crossing dynamics and phenomena are analogous to the super-horizon crossing of curvature perturbation modes during the inflation epoch, where they freeze outside the horizon and later reenter the horizon, influencing the formation of large-scale structures.
The article is organized as follows. In Sec.~\ref{review}, we briefly review the $X$ particles and $\bar X$ antiparticle productions and oscillations, and their dynamics effects on inflation \cite{Xue2023}, reheating \cite{ Xue2023a} and standard cosmology \cite{Xue2024} in the $\tilde\Lambda$CDM model. 
Section \ref{oscillationh} describes
the particle-antiparticle symmetric and asymmetric contrast density perturbations and 
derives their acoustic wave equations. In Sec.~\ref{mode0}, we 
analyze the wavelengths of acoustic modes. In
Secs.~\ref{ampath}, \ref{hcross0sub} and \ref{lcross} we discuss 
horizon crossings that result in particle and antiparticle asymmetry. 
Consequently, we calculate the baryon number-to-entropy ratio in Section~\ref{bary}. Additionally, building on preliminary studies of dark matter waves \cite{Xue2023}, Section~\ref{soundw} examines dark-matter acoustic waves for a given comoving wavelength and their relevance to physical observations and effects at large-distance scales. 
In this article, 
$G=M^{-2}_{\rm pl}$ is denoted as the the Newton constant, where $M_{\rm pl}$ is the Planck scale. The reduced Planck scale $m_{\rm pl}\equiv (8\pi)^{-1/2} M_{\rm pl}=2.43\times 10^{18} $GeV.

\comment{On the other hand, what is the crucial role that 
the cosmological $\Lambda$ term plays in inflation and reheating, and what is the essential reason for the coincidence of dark-matter dominant matter density and 
the cosmological $\Lambda$ energy density? 
There are various models and many efforts, 
that have been made to approach these issues, and readers are referred to 
review articles and professional books, for example, 
see Refs.~\cite{Peebles,kolb,book,Inflation_R,
Inflation_higgs,reviewL,Bamba:2012cp,Nojiri:2017ncd,
Coley2019,
prigogine1989,prigogine1989+,Khlopov,wangbin,martin,
wuyueliang2012,
wuyueliang2016,axioninf,xue2000,xuecos2009,xueNPB2015}.
} 

\section{\bf A brief review of $\tilde\Lambda$CDM applied to inflation and reheating}\label{review}

In the framework of the recently proposed $\tilde\Lambda$CDM ($\tilde\Lambda$ cold dark matter) scenario, we study dynamics and phenomena of inflation \cite{Xue2023},
reheating \cite{Xue2023a} and standard cosmology \cite{Xue2024} in accordance with observations.  
For the convenience of readers, we begin with a brief review of the main features of the $\tilde\Lambda$CDM scenario.

{\it First}, we propose a time-varying cosmological $\tilde\Lambda(t)$ term 
in the Friedman equation, representing dark energy interacting with matter and radiation. The Friedman equations for a flat Universe of horizon $H$ are \cite{Xue2015}
\begin{eqnarray}
H^2=\frac{8\pi G}{3}\rho;\quad \dot H=-\frac{8\pi G}{2}(\rho + p),\label{friedman}
\end{eqnarray}
where energy density $\rho\equiv \rho_{_M}+\rho_{_R}+\rho_{_\Lambda}$ and pressure 
$p\equiv p_{_M}+p_{_R}+p_{_\Lambda}$. The equations of state are  $p_{_{M,R,\Lambda}}= \omega_{_{M,R,\Lambda}} \rho_{_{M,R,\Lambda}}$, $\omega_{_{M}}=0$, $\omega_{_{R}}=1/3$ and $\omega_{_{\Lambda}}=-1$ for matter $\rho_{_M}$, radiation $\rho_{R}$ and dark energy $\rho_{_\Lambda}$ densities respectively.

The second equation of (\ref{friedman}) represents the generalized 
conservation law (Bianchi identity) for a time-varying cosmological term
(dart energy) $\rho_{_\Lambda}(t)\equiv \tilde\Lambda/(8\pi G)$, as it interacts with matter/radiation $\rho_{_{M,R}}$. It reduces to the usual equation 
$\dot \rho_{_{M,R}} + (1+\omega_{_{M,R}})H\rho_{_{M,R}}=0$ and $\dot\rho_{_\Lambda}=0$ when non-interacting $\rho_{_\Lambda}$ is constant in time. Future research will be on the microscopic origin and nature of the cosmological term and its interaction with matter.

\begin{figure*}[t]
\centering
\begin{center}
\includegraphics[height=5.5cm,width=9.8cm]{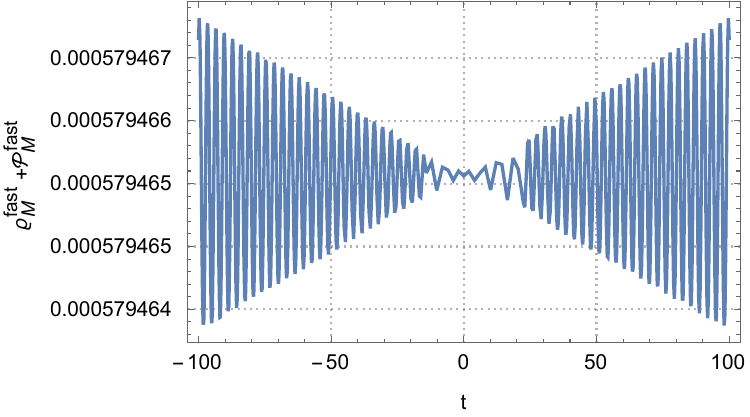}
\caption{We reproduce the Fig.~1 of Ref.~\cite{Xue2023a} to illustrate the coherent quantum oscillations on the state $|{\mathcal N}_{\rm pair}\rangle$, 
which are back-and-forth production and annihilation of massive particle and antiparticle pairs in microscopic space and 
time $t$ in the unit of the characteristic time scale $M^{-1}$ of quantum processes. It is described by quantum pair energy density $\varrho^{\rm fast}_{_M}$ and pressure ${\mathcal P}^{\rm fast}_{_M}$ (momentum density) 
in the unit of $\rho_{\rm crit}=3m_{\rm pl}^2H^2_{\rm slow}$, 
and the relevant dynamics equations are in Ref.~\cite{Xue2023a}. 
When the time $t\gg 1/M$, pair coherent oscillations
approach the equilibrium configuration described by a massive pair plasma fluid (\ref{apdenm}).} 
\label{reh-osci+}
\end{center}
\vspace{-2em}
\end{figure*}

{\it Second}, the massive particle-antiparticle pairs production \cite{Parker1973} and oscillation \cite{Xue2023, Xue2023a} establish a condensate ground state $|{\mathcal N}_{\rm pair}\rangle$ characterized by the large number (${\mathcal N}_{\rm pair}\gg 1$) of massive ($M\gg H$) pairs. These pairs are attributed to the microscopic
fast-component $H_{\rm fast}$ in the Hubble function $H=H_{\rm fast}+H_{\rm slow}$. The fast component $H_{\rm fast}$ oscillates coherently with massive pairs' quantum oscillation, which is related to the production, kinematic motion and annihilation of massive pairs at the ``microscopic'' 
spacetime scale $1/M$, as illustrated in Fig.~\ref{reh-osci+}. Massive particles and antiparticles are produced along with their kinetic energy and oscillate with $H_{\rm fast}$ in space and time due to their nonlinear interaction. These pair oscillations (fluctuations) imply that particles and antiparticles' local and instantaneous distributions (densities) are not identically in phase in space and time. However, particle and antiparticle symmetry conserves for the total particle and antiparticle numbers are the same. It is analogous to the dynamics and phenomena of creation and oscillation of electrons and positrons along with strong electric fields \cite{PhysRevLett.67.2427, Ruffini_2003,Ruffini_2010}.

The local and fast-oscillating $H_{\rm fast}$ dynamics behaviours are consistent with recent studies of vacuum fluctuation and ``microcyclic universes'' at small scales, demonstrated by local scale factor oscillation, as shown in Figure 1 of Refs.~\cite{Wang2020, Wang2020a}. 
The slow component $H_{\rm slow}$ obeys the Friedman equation 
(\ref{friedman}) at the ``macroscopic'' time scale $1/H$. The functional amplitude $H_{\rm slow}$ is much larger than $H_{\rm fast}$ with $H_{\rm slow}\approx H$, and subsequently, we will drop the subscript ``slow'' henceforth.

These ``microscopic'' and ``macroscopic'' processes are non-linearly coupled.
One has difficulties even numerically integrating coupled differential equations simultaneously due to the vast difference between the scales $1/M$ and $1/H$. Therefore, we have to model the condensate state $|{\mathcal N}_{\rm pair}\rangle$ as a macroscopic ``equilibrium'' or ``equipartition'' 
plasma fluid state by averaging the fast component $H_{\rm fast}$ and pairs' oscillations (Fig.~\ref{reh-osci+}) over microscopic time and length scales. 
This method allows us to study the back-reactions of fast component $H_{\rm fast}$ and pairs' oscillation state $|{\mathcal N}_{\rm pair}\rangle$ on the Friedman equation (\ref{friedman}). More detailed discussions are in Secs.~2-3, Figure 1 and Appendix ``Quantum pair oscillation details" of Ref.~\cite{Xue2023a}.

{\it Third}, we describe this ``equilibrium'' plasma fluid state as a perfect fluid state characterized by the effective number density $n^H_{_M}$ and energy density $\rho^H_{_M}$ of massive stable and unstable pairs of particles $X$ and antiparticles $\bar X$,
\begin{eqnarray}
\rho^H_{_M} \equiv  2\chi  m^2 H^2,\quad n^H_{_M} \equiv   \chi  m H^2, 
\label{apdenm}
\end{eqnarray}
the pressure and equation of state are $p^H_{_M}=\omega^H_{_M}\rho^H_{_M}$. Here, 
$m$ is the effective mass parameter representing the 
total mass and number of pairs in the massive pair plasma state. The lower limit $\omega^H_{_M}\approx 0$ for $m\gg H$ characterizes a non-relativistic massive plasma fluid in reheating \cite{Xue2023a}, which is relevant for the study in this article.  
Whereas, the upper limit $\omega^H_{_M}\lesssim 1/3$ for $m\gtrsim H$ 
is relevant for studying inflation \cite{Xue2023}.

Particle and antiparticle 
pair oscillations in spacetime establish the equilibrium state (\ref{apdenm}) when the time $t\gg 1/M$. 
This massive plasma fluid state is a holographic layer near the horizon because 
pair-oscillation contributions cancel each other from different patches inside the horizon. The spatial average of local pairs' and $H_{\rm fast}$ oscillations over the length scale $1/M$ vanish {\it inside} the horizon. 
The quantum oscillations of massive particle and antiparticle pairs' production, kinematic motion and annihilation occur near the horizon. We use the effective width parameter $\chi$ to characterize the layer radial width $\lambda_m = (\chi m)^{-1}\ll H^{-1}$.
We will adopt \footnote{In previous studies \cite{Xue2020, Xue2019}, a renormalization prescription 
at high energies $M\gg H$, which is different from the usual prescription (subtraction) at low energies $M\ll H$, consistently yields the mean density $n^H_{_M} \approx \chi m H^2$ (\ref{apdenm}) 
and effective width parameter $\chi\approx 1.85\times 10^{-3}$ by examining massive fermion pair productions in an exact De Sitter spacetime of constant $H$ and scaling factor $a(t)=e^{iHt}$. This result suggests $\chi\sim {\mathcal O}(10^{-3})$} $\chi=10^{-3}$ 
and treat $m$ as a free parameter, whose value depends on the Universe's evolution epoch.
Thus, at a given horizon $H$, the “macroscopic” condensation state of pair plasma fluid $\rho^H_{_M}$ (\ref{apdenm}) effectively represents the average overall ``microscopic'' states of the fast component $H_{\rm fast}$, pair production and oscillations at the time and length scale $1/M$. More detailed discussions are in Sec.~3-4 of Ref.~\cite{Xue2023a}.

\begin{figure*}[t]
\centering
\includegraphics[height=5.9cm,width=8.3cm]{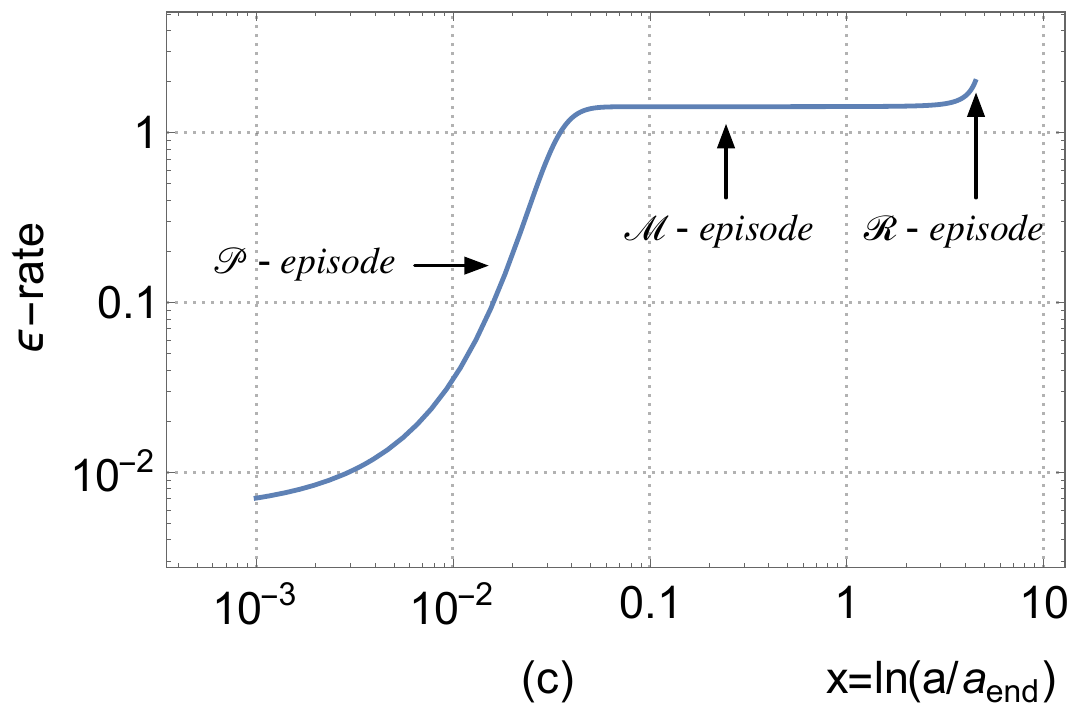}\hspace{0.333cm}
\caption{(Color Online). In a few $e$-folding number $x=\ln(a/a_{\rm end})$ during the reheating epoch, 
the $H$-variation $\epsilon$-rate (\ref{dde0}) indicates three distinct episodes: (i) $\epsilon < 1$ pre-reheating episode (${\mathcal P}$-{\it episode}) of massive pairs production, 
(ii) $\epsilon \approx 3/2$ massive pair oscillation domination 
episode (${\mathcal M}$-{\it episode}), and (iii) $\epsilon\approx 2$
radiation domination episode (${\mathcal R}$-{\it episode})
of genuine reheating due to massive pairs decaying into relativistic particles.
The ${\mathcal P}$-{\it episode} is much shorter than ${\mathcal M}$- and ${\mathcal R}$-{\it episodes}. The $a_{\rm end}$ indicates when the inflation ends $H\approx \Gamma_M$. The pre-inflation and inflation are in the regime $x=\ln(a/a_{\rm end})< 0$, where $H>\Gamma_M$ and $\epsilon \ll 1$, when dark-energy $\tilde \Lambda$ dominates. We reproduce this figure from Fig.~6 (c) of Ref.~\cite{Xue2023a}. 
The relevant dynamical equations are (\ref{friedman}), (\ref{rateeqd}) and (\ref{rateeqdr}). 
The time scale $H^{-1}$ governs this process, 
and time scale $\tau_{_M}=\Gamma^{-1}_M$ plays its role at the first turning point from $\mathcal P$ to $\mathcal M$,
and $\tau_{_R}=(\Gamma^{\rm de}_M)^{-1}$ plays its role at the second turning point from $\mathcal M$ to $\mathcal R$.
Note that in the ${\mathcal M}$- and ${\mathcal R}$-{\it episodes}, 
the evolution rate $\epsilon\approx 3/2, 2$ (\ref{dde0}) and the relaxation rate $\Gamma_M\propto \epsilon$  (\ref{prate}) are approximately constant in time.}
\label{AllBrR}
\end{figure*}

{\it Fourth}, we need to take into account how the horizon $H$ variation governed by the Freeman equations (\ref{friedman}) back-reacts on the equilibrium state (\ref{apdenm}), in other words, how the equilibrium state
responds to the $H$ variation. For this reason, 
we propose the cosmic rate equation (\ref{rateeqd}) to describe the interaction between the massive pair plasma energy density $\rho^H_{_M}$ (\ref{apdenm}) and normal matter density $\rho_{_{M}}$ in Freeman equation (\ref{friedman}), because their time scales of responding $H$ variation are very different. The normal matter density $\rho_{_{M}}$ time scale is $\tau_{_H}\approx H^{-1}$.
While, the massive pair plasma energy density $\rho^H_{_M}$ time scale $\tau_{_M}$ responding the $H$ variation can be estimated as follows,
\begin{eqnarray}
\tau^{-1}_{_M}=\Gamma_M, \quad  \Gamma_M &=& \frac{dN}{2\pi dt}\approx \frac{\chi m}{4\pi} \epsilon, \label{prate} 
\end{eqnarray}
where $N\approx n^H_{_M}H^{-3}/2$ is the pairs' number inside the Hubble sphere (averaged over volume)  
and $\Gamma_M$ is the rate of pair production and oscillation affected by the horizon $H$ change in time, governed by the Friedman equation (\ref{friedman}). We consider the $\tau_{_M}$ is a relaxation time of pair plasma fluid (\ref{apdenm}) responding to the $H$ variation.
This shows that the rate $\Gamma_M$ relates to the pair plasma fluid energy density $\rho^H_{_M}$ interaction with other energy density components $\rho_{_{R,M,\Lambda}}$ in the Friedman equation (\ref{friedman}).
The time scale $\tau_{_M}$ is significantly larger than the microscopic time scale $1/M$ for pairs' quantum oscillations (Fig.~\ref{reh-osci+}) but smaller than the macroscopic time scale $1/H$ of $\rho_{_{R,M,\Lambda}}$ density variation. It effectively describes the ``relaxation'' time scale, indicating how the massive pair plasma state varies in response to the macroscopic time as the Universe horizon $H(\rho_{_M},\rho_{_R},\rho_{_\Lambda})$ evolves.
The Universe evolution $\epsilon$-rate is usually defined as,
\begin{eqnarray}
\epsilon &\equiv& -\frac{\dot H}{H^2}=\frac{3}{2}\frac{(1+\omega_{_M})\rho_{_M}+ (1+\omega_{_R})\rho_{_R}}{\rho_{_\Lambda}+\rho_{_M}+\rho_{_R}}.
\label{dde0}
\end{eqnarray}
The second equation arises from the Friedman equations (\ref{friedman}). The asymptotic
values $\epsilon \ll 1$, $\epsilon \approx 2$ and $\epsilon \approx 3/2$ correspond to dark energy, radiation, and matter
domination, respectively. Fig.~\ref{AllBrR} shows the $\epsilon$ variation in the reheating, indicating the nonlinear behaviour of the crucial quantity $\Gamma_M\propto \epsilon$ (\ref{prate}) adopted for the analyses of this article.

The massive pair plasma state density $\rho^H_{_M}$ (\ref{apdenm}), associated with the horizon, contributes to the normal matter/radiation density $\rho_{_{M,R}}$. 
In turn,
the $\rho_{_{M,R}}$ variation affects the $\rho^H_{_M}$ via the horizon $H$. This interaction implies a non-linear back-and-forth
interplay between the massive pair plasma state and the matter state during
the Universe’s evolution. Moreover,
the massive pair plasma state has a
microscopic ``relaxation'' time scale $\tau_{_M}$ (\ref{prate}), differing from the macroscopic one $\tau_{_H}\approx H^{-1}$ of the matter and radiation state $\rho_{_{M,R}}$ in the Friedman equation (\ref{friedman}), i.e., $\tau_{_H}\gg \tau_{_M}$. Therefore, we cannot simply add $\rho^H_{_M}$ into $\rho_{_{M,R}}$ in the Friedman equation. 

By analogy with the Boltzmann rate equation for a microscopic back-and-forth process, e.g., $e^+e^-\Leftrightarrow\gamma\gamma$, in the macroscopic hydrodynamical expansion (see part of Eq.~(5.4) in  Ref.~\cite{Xue2023a}), we propose the back-and-forth interaction between the densities $\rho^H_{_M}$ and $\rho_{_{M,R}}$ follows the cosmic rate
equations of Boltzmann type, 
\begin{eqnarray}
\dot\rho_{_M}+ 3(1+\omega_{_M}) H\rho_{_M} &=& \Gamma_M(\rho_{_M}^H - \rho_{_M}-\rho_{_R}) - \Gamma_M^{^{\rm de}}\rho^{\rm de}_{_M},
\label{rateeqd}\\
\dot\rho_{_R}+ 3(1+\omega_{_R}) H\rho_{_R} &=& \Gamma_M(\rho_{_M}^H - \rho_{_M}-\rho_{_R}) + \Gamma_M^{^{\rm de}}\rho^{\rm de}_{_M}.\label{rateeqdr}
\end{eqnarray}
The term $3(1+\omega_{_{M,R}}) H\rho _{_{M,R}}$ with the time scale
$[3(1+\omega _{_{M,R}}) H]^{-1}$ represents the space-time expanding effect
on the density $\rho _{_{M,R}}$. The detailed balance term
$\Gamma _{M}(\rho _{_{M}}^{H} - \rho _{_{M}}- \rho _{_{R}})$ indicates how densities
$\rho _{_{M}}^{H}$ and $\rho _{_{M,R}}$ of different time scales couple together
in back-and-forth interaction. $\Gamma _{M} \rho _{_{M}}^{H}$ is the source
term, indicating $\rho _{_{M}}^{H}$ contribution to increasing
$\rho _{_{M,R}}$. $\Gamma _{M}\rho _{_{M,R}}$ is the depletion term, indicating
back reaction reducing $\rho _{_{M,R}}$. The ratio $\Gamma _{M}/H> 1$ indicates
the coupled case, and $\Gamma _{M}/H < 1$ indicates the decoupled case.

There are unstable massive pairs of density $\rho^{\rm de}_{_M}$ inside the massive pair plasma state (\ref{apdenm}). The term $-\Gamma^{^{\rm de}}_M\rho^{\rm de}_{_M}$ in 
(\ref{rateeqd}) represents unstable massive pairs decay to SM particles and other sterile dark-matter particles, whose masses are much smaller than massive pairs. They are ultra-relativistic particles (radiation) in the reheating epoch, described by the
cosmic rate equation (\ref{rateeqdr}) with the term $+\Gamma^{^{\rm de}}_M\rho^{\rm de}_{_M}$ for energy conservation. 
The decay rate is $\Gamma_M^{^{\rm de}}\approx g^2_{_Y}\hat m$ and time scale is
$\tau_{_R}=(\Gamma_M^{^{\rm de}})^{-1}$, where $g_{_Y}$ is 
the Yukawa coupling between massive pairs and relativistic particles.  
In the reheating epoch, we adopt the mass parameter 
$m=\hat m =27.7 m_{\rm pl}$ and $g^2_{_Y}= 10^{-9}$ \cite{Xue2023a}.
The cosmic rate equation (\ref{rateeqdr}) for relativistic particles 
reduces to the usual reheating equation \cite{Kolb1990} when $\Gamma^{^{\rm de}}_M\gg \Gamma_M$.  Actually, the 
scale $\tau_{_R}=(\Gamma_M^{^{\rm de}})^{-1}$ 
is the reheating time scale. 
The detailed discussions are in Secs.~4-5 of Ref.~\cite{Xue2023a}.   

To explicitly show the interaction between dark energy and matter, we can use the cosmic rate equations (\ref{rateeqd}) and (\ref{rateeqdr}) to recast the Friedman equations (\ref{friedman}) as a dark-energy equation
\begin{eqnarray}
\dot\rho_{_\Lambda}+3(1+\omega_{_\Lambda}) H\rho_{_\Lambda}&=& - 2 \Gamma_M \left(\rho^H_{_M} - \rho_{_{M}}-\rho_{_{R}}\right),
\label{rhoL}
\end{eqnarray}
where $\Gamma_M \left(\rho^H_{_M} - \rho_{_{M}}-\rho_{_{R}}\right)$ represents the interaction between dark energy and matter/radiation via the massive pair plasma state $\rho^H_{_M}$. This equation shows the dark energy interacting with matter and radiation via the non-linear 
back-and-forth balance term 
$\Gamma_M(\rho_{_M}^H - \rho_{_{M}}- \rho_{_{R}})$, the massive pair plasma state $\rho_{_M}^H$ (\ref{apdenm}), and the interacting rate $\Gamma_M$ (\ref{prate}), as shown in Fig.~\ref{AllBrR}.

During inflation and reheating epochs, the balance term $\Gamma_M \left(\rho^H_{_M} - \rho_{_{M}}-\rho_{_{R}}\right)$ is positive in Eqs.~(\ref{rateeqd},\ref{rateeqdr}) and (\ref{rhoL}). 
Therefore, $\dot\rho_{_\Lambda}<0$ and $\dot\rho_{_{M,R}}>0$, indicating that dark energy converts to matter and radiation energies \cite{Xue2023,Xue2023a}. In contrast, 
during standard
cosmology epoch after reheating, a negative 
balance term for $\rho^H_{_M}< \rho_{_{M}}+\rho_{_{R}}$ implies that matter and radiation convert to dark energy, resulting in $\dot\rho_{_\Lambda}> 0$ \cite{Xue2024}.

{\it In such $\tilde\Lambda$CDM scenario}, 
Eqs.~(\ref{friedman}-\ref{rateeqd}) form a close set of first-order
ordinary differential equations for the densities $\rho_{_{M,R,\Lambda}}$ and Hubble function $H$. The solutions are completely determined, provided observations fix initial conditions and effective mass parameters. We have investigated issues such as singularity-free and large-scale anomaly, the spectral index, and tensor-to-scalar ratio relation during the inflation epoch \cite{Xue2023}.
The Universe begins ($H>\Gamma_M$) with the dark-energy $\tilde\Lambda$ dominated pre-inflation ($\epsilon\gtrsim 0$) and inflation ($\epsilon\approx 0.0175$), which ends ($H\approx \Gamma_M$) and transitions to the reheating. In Ref.~\cite{Xue2023a}, particularly Secs.~7.2-7.3, we present detailed studies of the reheating epoch where the competition between the $\Gamma_M$ and $\Gamma_M^{^{\rm de}}$ rates plays an important role. As shown in Fig.~\ref{AllBrR}, the reheating starts ($\Gamma_M>H$) with the pre-reheating ${\mathcal P}$-episode ($\epsilon<1$), during which dark energy rapidly converts to massive pair production and oscillation.
Then, the ${\mathcal M}$-episode ($\epsilon\approx 3/2$) of massive pair domination occurs when $\Gamma_M >H> \Gamma_M^{^{\rm de}}$. This is followed by the genuine reheating ${\mathcal R}$-episode ($\epsilon\approx 2$) of radiation domination when $\Gamma_M^{^{\rm de}}> H>\Gamma_M$, during which unstable pairs decay to dark-matter particles and particles in the Standard Model for elementary particle physics. Additionally, we studied the cosmological
fine-tuning and coincidence problems in the standard cosmology \cite{Xue2024}.

In this article,  
we investigate the dynamics and phenomena of baryogenesis and dark-matter waves based on our studies and results during the reheating epoch \cite{Xue2023a}, where the plasma state mass parameter $m$ in Eqs.~(\ref{apdenm}) and (\ref{prate}) is specified as $\hat m$ and 
$\hat m=27.7 m_{\rm pl}$. First, we describe the
particle and antiparticle density perturbation of the plasma fluid state (\ref{apdenm}) in the layer near the horizon, then show how the horizon crossing dynamics of these perturbation modes relate to baryogenesis, dark-matter particle-antiparticle asymmetry and dark-matter waves.

\section
{\bf Particle and antiparticle density perturbations}\label{oscillationh}

The massive pair plasma fluid (\ref{apdenm}) is the ``semi-classical" equilibrium state containing the same number of particles $X$ and antiparticles $\bar X$ inside the horizon, and
the particle and antiparticle symmetry is preserved. However, their distributions (densities) $\rho^\pm_{_M}$ in spacetime are not identically in phase due to particle and antiparticle oscillations discussed in Sec.~\ref{review}. Thus, the densities $\rho^\pm_{_M}$ and relative 
(contrast) density $\rho^+_{_M}-\rho^-_{_M}$ do not vanish and possibly form density perturbations upon the equilibrium state (\ref{apdenm})   
when the horizon $H$ varies. The equilibrium state and these density perturbations have the relaxation time scale $\tau_{_M}=\Gamma_M^{-1}$ responding to the $H$ variation.

\subsection{Equations for particle and antiparticle density perturbations}\label{pertur}

To study their relative density contrast, we separate particles from antiparticles in the massive pair plasma state (\ref{apdenm}), and study their density perturbations respectively and examine symmetric and asymmetric density perturbations of massive particles and antiparticles. We use the notation $(+)$ for 
particles $X$  (matter) and $(-)$ for antiparticles $\bar X$ (antimatter). They have the same spin and mass $M\gg H$, but with opposite physical quantum numbers \footnote{The ``$\pm$'' do not indicate a positive and negative electric charge, we do not use ``$\dagger$'' for the sake of simplifying notations.}. They are not SM particles. However, some are unstable and decay or annihilate into SM particles and less massive dark matter particles during reheating, and stable ones remain as cold dark matter candidates.

\subsubsection{Continuity and Eulerian equations}

We focus on particle-antiparticle pairs' motion in the massive pair plasma layer near the horizon.
Analogously to the study of cosmic perturbation, see for example \cite{Kolb1990}, we 
separately study
the linear perturbations of massive particles and antiparticles by using their continuity equation and Eulerian equations of 
Newtonian motion of two perfect fluids of particle $(+)$ and 
antiparticle  $(-)$ 
densities $\rho^\pm_{_M}(t, {\bf x})$, pressures 
$p^\pm_{_M}(t, {\bf x})$ and velocities ${\bf v}^\pm_{_M}(t, {\bf x})$
in the Robertson-Walker space-time (Freemann Universe)
\begin{eqnarray}
&&\dot \rho^\pm_{_M} + {\bf \nabla}\cdot(\rho^\pm_{_M} {\bf v}^\pm_{_M})= \Gamma_M(\rho^0_{_M}-\rho^\pm_{_M})-\Gamma^{\rm de}_M\rho^\pm_{_M},  
\label{continuum}\\
&&\dot {\bf v}^\pm_{_M} + ({\bf v}^\pm_{_M}\cdot {\bf \nabla}){\bf v}^\pm_{_M} = - (\rho^\pm_{_M})^{-1}{\bf \nabla} p^\pm_{_M} - {\bf \nabla} \Phi,  \label{euler}\\
&&{\bf \nabla}^2\Phi = 4\pi G (\rho^+_{_M}+ \rho^-_{_M}),\label{newton}\\
&&\rho^0_{_M}=(1/2)\rho^H_{_M}= \chi\hat m^2 H^2, 
\label{rho0}
\end{eqnarray} 
and the last line $\rho^0_{_M}$ is the unperturbed mean density of equilibrium state. Note that we adopt non-relativistic equations (\ref{continuum}-\ref{euler}) for the dynamics of particle and antiparticle perfect fluids since they are very massive and non-relativistic. The densities $\rho^\pm_{_M}\sim p^H_{_M}$ and 
pressures $p^\pm_{_M}\sim p^H_{_M}$ are components of the massive plasma fluid (\ref{apdenm}) with an equation of state $p^H_{_M}=\omega^H_{_M}\rho^H_{_M}$ and $\omega^H_{_M}\ll 1$. 
These equations are valid only inside the horizon, where the causality for these equations holds. The characteristic time scale of perturbation equations (\ref{continuum}) and (\ref{euler}) relates to the relaxation time scale $\tau_{_M}=\Gamma^{-1}_M$. 

In Eq.~(\ref{continuum}), we use the detailed balance term $\Gamma_M(\rho^0_{_M}-\rho^\pm_{_M})$ to describe fluctuating densities $\rho^\pm_{_M}$ around the mean density with the time scale $\tau_{_M}=\Gamma^{-1}_M$. 
The term $\Gamma^{\rm de}_M\rho_{_M}$ describes unstable massive pair decay of the time scale $\tau_{_R}=(\Gamma^{\rm de}_M)^{-1}$. The arguments in the above equations are comoving coordinates $({\bf x}, {\bf k})$ and the time derivative and space gradients are taken with respect to the physical coordinates $(a{\bf x}, {\bf k}/a)$, where $a$ is the scale factor. The zeroth order solutions 
$\rho^\pm_{_M}$ to Eqs.~(\ref{continuum},\ref{euler},\ref{newton}) 
are the mean density $\rho^0_{_M}$ of massive pair plasma, 
which follows the Hubble flow 
${\bf v}^0_{_M}=d(a{\bf x})/dt =H a {\bf x}$, and gives 
the gravitational potential $\Phi^0=(2\pi G/3)\rho^0_{_M} |{\bf x}|^2$.

The interacting rates $\Gamma_M$ and $\Gamma^{\rm de}_M$, the gravitational potential $\Phi$ and mean density $\rho^0_{_M}=(1/2)\rho^H_{_M}$ are invariant under 
the particle and antiparticle transformation. Namely,
these equations are symmetric for $\rho^+_{_M}\leftrightarrow \rho^-_{_M}$. Therefore, Equations (\ref{continuum}-\ref{rho0}) fully respect 
the symmetry of particle and antiparticle. 
We will neglect the decay term $\Gamma^{\rm de}_M\rho_{_M}$ by considering $\Gamma_M\gg \Gamma^{\rm de}_M$, namely, the massive pair
plasma density variation rate is much larger than the unstable pair decay rate. 
The argument for neglecting $\Gamma^{\rm de}_M$ will be given later in Fig.~\ref{crossingf} caption. 

In the density perturbation equations (\ref{continuum}-\ref{rho0}), the only new term is the $\Gamma_M(\rho^0_{_M}-\rho^\pm_{_M})$, compared with corresponding canonical equations in literature. This term indicates the $\rho^\pm_{_M}$ fluctuations upon $\rho^0_{_M}$. These fluctuations are caused by the horizon $H$ variation in time, and the rate $\Gamma_M$ quantifies their response to the horizon $H$ variation. One should distinguish the continuum equation (\ref{continuum}) from the cosmic rate equation (\ref{rateeqd}). The former is the equation for particle (antiparticle) density perturbations inside the massive pair plasma. The latter is the equation for the massive pair plasma density interacting with the matter and dark energy densities in the Friedman equation.

\subsubsection{Contrast density perturbations}

Here the usual approach of density perturbation is adopted for analysis. We consider the small perturbations around the averaged mean values $\rho^0_{_M}$ and ${\bf v}^0_{_M}$ of massive pair plasma by writing
\begin{eqnarray}
\delta {\bf v}^\pm_{_M}={\bf v}^\pm_{_M}-{\bf v}^0_{_M},\quad 
\delta\rho^\pm_{_M} = \rho^\pm_{_M}-\rho^0_{_M}\quad {\rm and}\quad \delta^\pm_{_M}=\delta\rho^\pm_{_M}/\rho^0_{_M}. 
\label{flu}
\end{eqnarray}
They are functions of spacetime $({\bf x},t)$. Up to the first order in the perturbative quantities, 
the mean density $\rho^0_{_M}=\rho^{H}_{_M}/2$ is approximated as a constant in spacetime, and Equations (\ref{continuum}), (\ref{euler}) and (\ref{newton}) become
\begin{eqnarray}
&& d(\delta \rho^\pm_{_M})/dt + \rho^0_{_M}{\bf \nabla}\cdot(\delta{\bf v}^\pm_{_M})+ 3H \delta \rho^\pm_{_M} = -\Gamma_M\delta\rho^\pm_{_M},  
\label{pcontinuum}\\
&&d (\delta{\bf v}^\pm_{_M})/dt + H\delta {\bf v}^\pm_{_M} = - (\rho^0_{_M})^{-1}{\bf \nabla} \delta p^\pm_{_M} - {\bf \nabla} \delta\Phi,  \label{peuler}
\end{eqnarray}
and the Poisson equation becomes
\begin{eqnarray}
{\bf \nabla}^2\delta \Phi = 4\pi G (\delta\rho^+_{_M}+ \delta\rho^-_{_M}). 
\label{poi}
\end{eqnarray}
In terms of $\delta^\pm_{_M}=\delta\rho^\pm_{_M}/\rho^0_{_M}$,
Equation (\ref{pcontinuum}) yields 
\begin{eqnarray}
{\bf \nabla}\cdot \delta {\bf v}^\pm_{_M}=-(\dot\delta^\pm_{_M}  + \Gamma_M \delta^\pm_{_M}). 
\label{vpm}
\end{eqnarray}
Taking the gradient of Eq.~(\ref{peuler}), we arrive at
\begin{eqnarray}
\ddot\delta^\pm_{_M} +(\Gamma_M+ 2H)\dot\delta^\pm_{_M} 
+(2H\Gamma_M +\dot\Gamma_M) \delta^\pm_{_M} &=& v_s^2{\bf \nabla}^2 \delta^\pm_{_M} 
+ 4\pi G\rho^0_{_M} (\delta^+_{_M}+ \delta^-_{_M}) 
\label{dpeuler}
\end{eqnarray}
where the sound velocity 
\begin{eqnarray}
v_s=(\delta p^\pm_{_M}/\delta \rho^\pm_{_M})^{1/2}
\label{soundv}
\end{eqnarray} 
can be obtained from the equation of state below Eq.~(\ref{apdenm}).
Since particles and antiparticles are massive, their fluid sound velocity $v_s\ll 1$ and $\delta^\pm_{_M}$ are 
non-relativistic density perturbations. Equation (\ref{dpeuler}) represents the perturbations 
of particle and antiparticle densities. It reduces to the usual equation for density perturbation in the case $\Gamma_M =0$ and $\dot\Gamma_M =0$. We note that (i)
particle and antiparticle contrast density perturbations are treated separately; (ii) particle and antiparticle symmetry is preserved in these second-order partial differential equations at microscopic scales in time and space inside the horizon.

\subsection{Equations for symmetric and asymmetric density perturbations}
\label{Asypert}

To describe the perturbation of the 
symmetric particle-antiparticle pair density, and
the perturbation of the asymmetrical particle-antiparticle density, we introduce density contrasts:
\begin{eqnarray}
\Delta_{_M} &\equiv & (\delta^+_{_M} + \delta^-_{_M})/2=(\rho^{+-}_{_M}-\rho^H_{_M})/\rho^H_{_M},
\label{delta1}\\
\delta_{_M} &\equiv & (\delta^+_{_M} - \delta^-_{_M})/2=(\rho^+_{_M}-\rho^-_{_M})/\rho^H_{_M},
\label{delta2}
\end{eqnarray}
where the sum $\rho^{+-}_{_M}\equiv \rho^+_{_M}+\rho^-_{_M}$. 
Henceforth, we call $\Delta_{_M}$ the pair-density $\rho^{+-}_{_M}$ perturbation and 
$\delta_{_M}$ the particle-antiparticle-density perturbation. The $\Delta_{_M}$ is the density contrast between perturbed pair density $\rho^{+-}_{_M}$ and unperturbed mean density $\rho^H_{_M}$. The $\delta_{_M}$ is the density contrast between particles' density $\rho^+_{_M}$ and antiparticles' density $\rho^-_{_M}$. 
The former is symmetric, whereas the latter is antisymmetric under the particle and antiparticle transformation $\rho^+_{_M}\leftrightarrow \rho^-_{_M}$. The perturbations $\Delta_{_M}$ and $\delta_{_M}$ vanish, when both $\rho^+_{_M}$ and $\rho^-_{_M}$ approach to the mean value $\rho^0_{_M}=(1/2)\rho^H_{_M}$. 

The vanishing perturbation $\delta_{_M}= 0$ indicates 
that particle and antiparticle distributions are identical in spacetime, e.g., they are in phase. The vanishing $\Delta_{_M}= 0$ indicates no density perturbation of massive plasma fluid (\ref{apdenm}) in spacetime. Instead, the 
non-vanishing $\delta_{_M}\not = 0$ means 
that particle and antiparticle distributions are not in phase in spacetime, implying the different spatial distributions of particles and antiparticles that change in time. The non-vanishing $\Delta_{_M}\not=0$
implies nontrivial density perturbation upon the equilibrium state $\rho^H_{_M}$ (\ref{apdenm}). 
Here, we stress that the perturbations $\delta_{_M}$ and $\Delta_{_M}$ are not only functions of space points inside the horizon but also functions of time connecting with the horizon $H$ evolution. 

\subsubsection{Acoustic wave  equations}

Replacing Eqs.~(\ref{delta1},\ref{delta2}) in Eq.~(\ref{dpeuler}), we obtain
acoustic wave  equations for the asymmetric $\delta_{_M}$  and symmetric  $\Delta_{_M}$ perturbations,
\begin{eqnarray}
\ddot\delta_{_M} +(\Gamma_M+ 2H)\dot\delta_{_M} 
+(2H\Gamma_M +\dot \Gamma_M)\delta_{_M} &=& v_s^2{\bf \nabla}^2 \delta_{_M}, 
\label{dp+}\\
\ddot\Delta_{_M} +(\Gamma_M+ 2H)\dot\Delta_{_M} 
+(2H\Gamma_M +\dot \Gamma_M) \Delta_{_M} &=& v_s^2{\bf \nabla}^2 \Delta_{_M} 
+ 4\pi G\rho^H_{_M} \Delta_{_M}.
\label{dp-}
\end{eqnarray}
It is shown that the modes $\Delta_{_M}$ and $\delta_{_M}$ satisfy the same type of oscillating equation, except an additional term $4\pi G\rho^H_{_M} \Delta_{_M}$ 
in Eq.~(\ref{dp-}) due to massive pairs in the external gravitational potential. Equations (\ref{dp+}) and (\ref{dp-}) are wave ($v_s\not= 1$) equation or oscillating ($v_s=0$) equations for non-relativistic density contrast perturbations. In the inflation epoch 
and the ${\mathcal M}$-episode of the reheating, the $\epsilon$-rate (\ref{dde0}) varies slowly in time, see Fig.~\ref{AllBrR}. We will neglect  $\dot\Gamma_M=(\chi \hat m/4\pi)\dot\epsilon\gtrsim 0$. 

These second-order partial differential equations (\ref{dp+}) and (\ref{dp-}) are valid inside the horizon and respect the asymmetry and symmetry under particle and antiparticle transformation $\rho^+_{_M}\leftrightarrow \rho^-_{_M}$.
They are homogeneous and linear in perturbations $\delta_{_M}$ and $\Delta_{_M}$  without external source terms. Their nontrivial solutions depend only on (i) the ``initial'' values of $(\delta_{_M},\Delta_{_M})$ and 
$(\dot \delta_{_M},\dot \Delta_{_M})$; (ii) the boundary values of $(\delta_{_M},\Delta_{_M})$ and gradients $(\nabla\delta_{_M},\nabla\Delta_{_M})$ on the horizon surface. The former can be created (triggered) by quantum pair oscillations shown in Fig.~\ref{reh-osci+}, otherwise $\delta_{_M}({\bf x},t)\equiv 0$ and $\Delta_{_M}({\bf x},t)\equiv 0$ identically vanish in spacetime.


We define the Fourier transformation from $f_{_M}({\bf x},t)=\Delta_{_M}({\bf x},t),\delta_{_M}({\bf x},t)$ to ${\bf k}$ 
modes 
$f^{\bf k}_{_M}(t)=\Delta^{\bf k}_{_M}(t),\delta^{\bf k}_{_M}(t)$,
\begin{eqnarray}
f_{_M}({\bf x},t) &=& \frac{1}{V^{1/2}}\sum_{\bf k}f^{\bf k}_{_M}(t)e^{ i{\bf k}{\bf x}},
\label{fm1}\\
f^{\bf k}_{_M}(t) &=& \frac{1}{V^{1/2}}\int d^3x f_{_M}({\bf x},t)e^{-i{\bf k}{\bf x}},
\label{fm2}
\end{eqnarray} 
where $V=(4\pi/3)H^{-3}$ is the physical Hubble volume. 
The corresponding wave-propagating equations for ${\bf k}$-modes' 
$\delta^{\bf k}_{_M}$ and $\Delta^{\bf k}_{_M}$ are approximately given by, 
\begin{eqnarray}
\ddot\delta^{\bf k}_{_M} +(\Gamma_M+ 2H)\dot\delta^{\bf k}_{_M} 
+2H\Gamma_M \delta^{\bf k}_{_M} +(v_s^2|{\bf k}|^2/a^2) \delta^{\bf k}_{_M}&=& 0
\label{kdp+}\\
\ddot\Delta^{\bf k}_{_M} +(\Gamma_M+ 2H)\dot\Delta^{\bf k}_{_M} 
+2H\Gamma_M\Delta^{\bf k}_{_M} + (v_s^2|{\bf k}|^2/a^2) \Delta^{\bf k}_{_M} 
&=&  4\pi G\rho^H_{_M} \Delta^{\bf k}_{_M}.
\label{kdp-}
\end{eqnarray}
These are semi-classical equations governing the oscillating modes
$\delta^{\bf k}_{_M}$ and $\Delta^{\bf k}_{_M}$ of effective 
frequencies
\begin{eqnarray}
\omega_\delta^2({\bf k})&=&2H\Gamma_M  + (v_s^2|{\bf k}|^2/a^2) 
\label{fdp+}\\
\omega_\Delta^2({\bf k})&=&2H\Gamma_M  + (v_s^2|{\bf k}|^2/a^2) - 4\pi G\rho^H_{_M}.
\label{fdp-}
\end{eqnarray}  
The $\Gamma_M$-term contributes to the friction coefficient $(\Gamma_M+2H)$ in oscillation equations 
and ``quasi mass" term $(2H\Gamma_M)^{1/2}$ in the oscillation frequency 
$\omega_{\delta,\Delta}^2({\bf k})$. 
The fast oscillating perturbation modes $\delta({\bf k})$  
and $\Delta({\bf k})$ have a small time scale $\omega_{\delta,\Delta}^{-1}({\bf k})\leq (2H\Gamma_M)^{-1/2}$. Here, we consider an adiabatic approximation of $H\Gamma_M$ 
being almost constant on such a small time scale. 
The reason is that $H$ and $\Gamma_M$ are slowly time-varying functions governed by the Freedman equation (\ref{friedman}) and cosmic rate equations (\ref{rateeqd},\ref{rateeqdr}) in macroscopic time scales $\tau_{_H}$ and $\tau_{_M}$, see the reheating ${\mathcal M}$-{\it episode} in Fig.~\ref{AllBrR}.

The term  $4\pi G\rho^H_{_M}$ in Eqs.~(\ref{kdp-},\ref{fdp-}) 
could lead to the Jeans instability, due to the gravitational attraction of massive pair plasma, possibly 
giving a nontrivial solution $\Delta_{_M}$. 
Observe that in Eq.~(\ref{fdp-}) 
$4\pi G\rho^H_{_M}=8\pi\chi ( \hat m/M_{\rm pl})^2 H^2$ is much smaller than $2H\Gamma_M + (v_s^2|{\bf k}|^2/a^2)$ 
even for the case $|{\bf k}|=0$ and $\hat m\gg H$ ($v_s^2\ll 1$)
in the inflation and reheating epochs, as well as standard cosmology. 
Therefore the negative term $4\pi G\rho^H_{_M}$ can be neglected 
and $\omega_\Delta^2({\bf k})>0$, implying the Jeans instability should not occur in these epochs.

\subsubsection{Lowest-lying mode oscillating equations}

We will discuss the solutions to Eqs.~(\ref{kdp+}-\ref{fdp-}) for the density perturbations in the inflation epoch and three episodes ${\mathcal P}$, ${\mathcal M}$ and ${\mathcal R}$ of the reheating epoch, see Fig.~\ref{AllBrR}. 
We first focus on the lowest-lying oscillation modes by neglecting the pressure term 
$v_s^2{\bf \nabla}^2$ or $(v_s^2|{\bf k}|^2/a^2)$ terms for wave propagation, 
since massive pairs' plasma fluid (\ref{apdenm}) 
is non-relativistic  
and their sound velocity is small $v_s^2 \ll 1$. Equations 
(\ref{kdp+}-\ref{fdp-}) become oscillating equations,
\begin{eqnarray}
\ddot\delta^0_{_M} +(\Gamma_M+ 2H)\dot\delta^0_{_M} 
+2H\Gamma_M \delta^0_{_M} &=& 0, 
\label{kdp+0}\\
\ddot\Delta^0_{_M} +(\Gamma_M+ 2H)\dot\Delta^0_{_M} 
+2H\Gamma_M\Delta^0_{_M} &=& 0,
\label{kdp-0}
\end{eqnarray}
where the lowest lying oscillation modes $\delta^0_{_M}\equiv \delta_{_M}^{{\bf k}={\bf 0}}$ and $\Delta^0_{_M}\equiv \Delta_{_M}^{{\bf k}={\bf 0}}$ with the frequencies $\omega_\delta\equiv \omega_\delta({|\bf k|=0})$ and $\omega_\Delta\equiv \omega_\Delta({|\bf k|=0})$,
\begin{eqnarray}
\omega_\delta^2&=&\omega_\Delta^2=2H\Gamma_M.
\label{fdp-0}
\end{eqnarray}
We call the lowest-lying modes as ``zero modes'' $\Delta^0_{_M}$ and 
$\delta^0_{_M}$ of pair-density and particle-antiparticle-density oscillations. In this semi-classical approximation, these zero modes' frequency (\ref{fdp-0}) 
weakly depend on the time $t$ in the period of slowly time-varying $H$ and $\Gamma_M$.  

We note that Eqs.~(\ref{kdp+0}) and (\ref{kdp-0}) have the same structure, and they are no longer oscillating equations when $\Gamma_M=0$, namely, microscopic pair oscillations decouple from the macroscopic horizon evolution. In this case, the solutions damp in time, $\Delta^0_{_M}\rightarrow 0$ and $\delta^0_{_M}\rightarrow 0$. Namely, $\rho_{_M}^+,\rho_{_M}^- \rightarrow \rho_{_M}^0$ and $\rho_{_M}^{+-}\rightarrow \rho_{_M}^H$, see Eqs.~(\ref{delta1}) and (\ref{delta2}),
density perturbations do not occur. As will be seen, $\Gamma_M\not=0$ is crucial for discussing density perturbation dynamics in this article.

To end this section, we would like to mention that the oscillations $\delta^0_{_M}$
are the spatial fluctuations in the number of particles or antiparticles (compositions) 
per comoving volume, and the oscillations $\Delta^0_{_M}$ are the spatial fluctuations in the number of pairs per comoving volume.  

\subsubsection{Particle-antiparticle asymmetry due to modes' horizon crossing}\label{reason}

Before analyzing acoustic wave equations (\ref{dp+}) and (\ref{dp-}) in the coordinate space $\bf{x}$ or 
their counterparts (\ref{kdp+}) and (\ref{kdp-}) in the momentum space $\bf{k}$, we discuss in general the possible situations for nontrivial particle-antiparticle asymmetric solutions inside the horizon due to modes' horizon crossing. 

The local and instantaneous perturbations of relative density $\delta_{_M}({\bf x},t)\not=0$ produced by quantum pair oscillations (Fig.~\ref{reh-osci+}) is necessary but not sufficient for creating particle and antiparticle asymmetry, namely net particle numbers observed inside the horizon.
We need to examine the total net particle numbers $\int_V d^3x \rho^{H}_{_M}\delta_{_M}=\hat m(N^+-N^-)\equiv D_{+-}$, where $N^\pm$ are the total numbers of particles and antiparticles inside the horizon. Integrating Eq.~(\ref{dp+}) overall the horizon volume $V$, we obtain inside the horizon,
\begin{eqnarray}
\ddot D_{+-} &+& (\Gamma_M+ 2H)\dot D_{+-} 
+(2H\Gamma_M +\dot \Gamma_M)D_{+-} \nonumber\\
&=& v_s^2\int_V d^3x{\bf \nabla}^2 (\delta_{_M}\rho^{H}_{_M})= v_s^2\int_{\partial V} d{\bf S}\cdot {\bf \nabla} (\delta_{_M}\rho^{H}_{_M}), 
\label{Ndp+}
\end{eqnarray}
where the right-handed side for ${\bf \nabla}(\delta_{_M}\rho^{H}_{_M})|_{\partial V}\not=0
$ and $\delta_{_M}|_{\partial V}\not=0$ represents the difference of particles and antiparticles crossing in or out the horizon surface $\partial V$ of the 
horizon volume $V$. 

In our framework, there is no any nontrivial ``initial'' values $\dot D^0_{+-}\not=0$ and  
$D^0_{+-}\not=0$ caused by the external sources of particle and antiparticle asymmetric interactions and processes because of no explicit or spontaneous breaking of particle and antiparticle symmetry in microscopic interactions. However, there are the local relative (contrast) densities $\delta^{\bf k}_{_M}\not=0$, which are caused by quantum pair oscillations (Fig.~\ref{reh-osci+}) of particle and antiparticle production, kinetic motion and annihilation inside the horizon.

Thus, in this framework
``initial'' values $\dot D^0_{+-}=0$ and  
$D^0_{+-}=0$ of particles and antiparticles vanish 
inside the horizon, 
but their local relative contrast perturbations $\delta_{_M}({\bf x},t)\not=0$ are nontrivial in spacetime. 
We discuss the possibilities of
nontrivial asymmetric solution $\dot D_{+-}(t)\not=0$ and 
$D_{+-}(t)\not=0$ appearing inside the horizon as the horizon $H$ evolves in time:
\begin{enumerate}[(a)]
\item the trivial boundary values ${\bf \nabla} (\delta_{_M}\rho^{H}_{_M})|_{\partial 
V}=0$ and $\delta_{_M}|_{\partial V}=0$, representing there is no particle-antiparticle horizon crossing. It means in momentum mode space (\ref{kdp+}), all $\delta^{\bf k}_{_M}$ modes' wavelength $\omega_\delta^{-1}({\bf k})$ 
(\ref{fdp+}) are smaller than the horizon size $H^{-1}$, i.e., $\omega_\delta({\bf k})>H$. All modes $\delta^{\bf k}_{_M}$ are inside the horizon. As a result of linear and homogeneous Eq.~(\ref{Ndp+}), 
$\dot D_{+-}(t)=0$ and $D_{+-}(t)=0$, no particle-antiparticle asymmetry occurs inside the horizon of the volume $V$, which contains all particles and antiparticles $N^+\equiv N^-$.
\item the nontrivial boundary values ${\bf \nabla} (\delta_{_M}\rho^{H}_{_M})|_{\partial 
V}\not=0$ and $\delta_{_M}|_{\partial V}\not=0$, representing the particle-antiparticle horizon 
crossing. It means in momentum mode space (\ref{kdp+}) some $\delta^{\bf k}_{_M}$ modes' wavelength $\omega_\delta^{-1}({\bf k})$ (\ref{fdp+}) 
exceeds the horizon size $H^{-1}$, i.e., $\omega_\delta({\bf k})<H$, thus they cross outside the horizon. Nontrivial boundary conditions or non-vanishing 
right-handed side of Eq.~(\ref{Ndp+}) implies that 
$\dot D_{+-}(t)\not=0$ and $D_{+-}(t)\not=0$ and particle-antiparticle asymmetry can possibly occur inside the horizon. However, the total number of particles inside and outside the horizon conserves,
\begin{eqnarray}
\int_V d^3x \rho^{H}_{_M}\delta_{_M}+ \int_{V_{\rm out}} d^3x \rho^{H}_{_M}\delta_{_M}\equiv 0, 
\label{totalN}
\end{eqnarray}
where the second term $\int_{V_{\rm out}} d^3x \rho^{H}_{_M}\delta_{_M}=D^{\rm out}_{+-}=\hat m(N^+_{\rm out}-N^-_{\rm out})$ represents the difference between particles and antiparticles outside the horizon $V_{\rm out}$. The sub-horizon observers see 
the particle-antiparticle asymmetry $D_{+-}=-D^{\rm out}_{+-}\not=0$, whose value relates to crossing-mode $\delta_{_M}|_{\omega_\delta({\bf k})=H}$ amplitudes on the horizon surface $\partial V$ 
at the horizon-crossing moment. The horizon boundary term on 
the right-handed side 
of Eq.~(\ref{Ndp+}) is proportional to 
\begin{eqnarray}
4\pi H^{-2}\rho^{H}_{_M}H\delta_{_M}\big|_{\partial V}= 8\pi \chi \hat m^2H\delta_{_M}\big|_{\partial V}, 
\label{pV}
\end{eqnarray}
which is nontrivial as long as 
$\delta_{_M}|_{\partial V}\not=0$.
\end{enumerate}
Similar discussions also apply to Eq.~(\ref{dp-}) for the density perturbation $\Delta_{_M}$, which we omit here to simplify the text. The above discussions are essentially 
same as those, see below Eqs.~(\ref{dp+}) and (\ref{dp-})
on initial and boundary conditions necessary for nontrivial solutions to second-order partial 
differential equations.  
We thus turn to the discussions of the horizon-crossing possibility (b) of the difference between particle and antiparticle modes crossing the horizon surface in the following two distinct cases. 

First, in the case of horizon $H={\rm const}$ and volume $V$ fixed in time and $\Gamma_M=0$, Eqs.~(\ref{dp+}) and (\ref{kdp+}) become usual equations for perturbations 
\begin{eqnarray}
\ddot\delta_{_M} +2H\dot\delta_{_M} -
v_s^2{\bf \nabla}^2 \delta_{_M} &=& 0.
\label{dp+0}
\end{eqnarray}
The particle and antiparticle symmetry $D_{+-}=0$ 
holds inside the horizon, assuming 
particle-antiparticle oscillations $\delta_{_M}$ 
are confined inside the horizon $\delta_{_M}|_{\partial V}=0$ and ${\bf \nabla} (\delta_{_M}\rho^{H}_{_M})|_{\partial V}=0$. All sub-horizon modes $\delta^{\bf k}_{_M}$ 
obey the wave-mode equation 
\begin{eqnarray}
\ddot\delta^{\bf k}_{_M} +2H\dot\delta^{\bf k}_{_M} +(v_s^2|{\bf k}|^2/a^2) \delta^{\bf k}_{_M} 
&=& 0,
\label{kdp+00}
\end{eqnarray}
with physical wavelength $\omega_{\delta,\Delta}^{-1}({\bf k})=(v_s|{\bf k}|/a)^{-1}< H^{-1}$. 
This sub-horizon condition
does not change in time, which is the crucial point.
Therefore, the particle and antiparticle symmetry $D_{+-}(t)=0$ preserves in time. Moreover, the relative density perturbations $\delta_{_M}$ or modes $\delta^{\bf k}_{_M}$ inside the horizon 
$\delta_{_M},\delta^{\bf k}_{_M}\propto e^{-Ht}$ dampen down in time. The similar discussion applies to Eq.~(\ref{kdp-0}) for the density perturbations $\Delta^{\bf k}_{_M}$, if we neglect the small gravitational attraction term $4\pi G\rho^H_{_M} \Delta^{\bf k}_{_M}$. 

Second, we come to the case that slowly time-varying $H$ and $\Gamma_M\not=0$ are present in adiabatically approximated wave-mode equations (\ref{kdp+}) and (\ref{kdp-}). The effective frequencies (\ref{fdp+}) and (\ref{fdp-}) are time dependent via the quasi mass term $2H\Gamma_M$. Therefore, 
the sub-horizon condition $\omega_{\delta}^{-1}({\bf k})< H^{-1}$
can change to the super-horizon condition $\omega_{\delta}^{-1}({\bf k})> H^{-1}$ 
in time, {\it vice versa}. Thus there is a possibility in time evolution 
that the symmetric $D_{+-}=0$ case of 
{\it all} particle and antiparticle oscillating modes inside the horizon (sub-horizon) can change to the asymmetric $D_{+-}\not=0$ case of some modes outside the horizon (super-horizon) at the horizon crossing $\omega_{\delta}^{-1}({\bf k})= H^{-1}$, {\it vice versa}. Namely, as the horizon $H$ and rate $\Gamma_M$ evolve in time, the aforementioned sub-horizon case (a) and the super-horizon case (b)  exchange. As a consequence, some particles or antiparticles cross in or out the horizon surface $\partial V$ and ${\bf \nabla} (\delta_{_M}\rho^{H}_{_M})|_{\partial V}\not=0$ 
in Eq.~(\ref{Ndp+}).

In other words, due to the 
horizon-expanding and spacetime strengthening, particles and antiparticles decouple from oscillating each other, causing the possibility that particles remain inside the horizon and antiparticles go outside the horizon. Particles or antiparticles that cross outside the horizon are out of causality, and their oscillating amplitudes $\delta^{\bf k}_{_M}$ are frozen, and the values depend on when the horizon-crossing occurs. On the contrary, in time evolution, there is also the possibility that antiparticle (particle) super-horizon modes return to the horizon and couple to particle (antiparticle) modes, undergoing sub-horizon pair oscillation modes.

Moreover, the massive pair plasma fluid (\ref{apdenm}) 
is an holographic layer near the horizon $H$, which evolves on the time scale $\tau_{_H}=H^{-1}\gg \tau_{_M}$. 
During the horizon evolution, the particle and antiparticle relative density perturbation modes can cross the horizon, depending on their wavelength compared with the horizon size. It is a dynamic of competition between the horizon-expanding rate $H$ and the particle-antiparticle oscillation rate $\omega_{\delta,\Delta}({\bf k})$. In a semi-classical picture, this dynamical phenomenon illustrates
when spacetime expands so fast that particle and antiparticle pairs do not have enough time to proceed 
with their back-and-forth oscillations, thus 
decoupling from each other. Therefore, particles (or antiparticles) 
go out of the horizon, and their partners remain inside the horizon. Similar dynamics are studied in the neutrinos decoupling \cite{Lee1977} in the early Universe and electrons and positrons decoupling in their pair plasma 
undergoing an ultra-relativistic hydrodynamical expansion \cite{1999A&A...350..334R, Ruffini1999b, 2000A&A...359..855R}. 

In the next sections, we attempt to study such horizon-crossing dynamics of massive pair plasma oscillations, i.e., the sub-horizon case (a) and super-horizon case (b), in the reheating epoch to see possible physical consequences.

\comment{Equation (\ref{kdp+0}) is homogeneous and linear in $\delta^{\bf k}_{_M}$, nontrivial solutions $\delta^{\bf k}_{_M}(t)\not=0$ require a non-vanishing value $\delta^{\bf k}_{_M}(t_0)\not=0$ at an ``initial'' time $t_0$. This value can be caused by quantum fluctuations (oscillations) (Figs.~\ref{reh-osci+}) of particle and antiparticle production, kinetic motion, and annihilation inside the horizon, as discussed in Sec.~\ref{review}. This does not mean the particle and antiparticle symmetry,  
because particle numbers $N^+$ and antiparticle numbers $N^-$ are the same inside the horizon. Using Eq.~(\ref{fm1}) and $V^{-1}\int_Vd^3x e^{i{\bf k}{\bf x}}=\delta_{{\bf k},0}$, from 
$\int_Vd^3x \delta^{\bf k}_{_M}\propto (N^+-N^-)=0$, we obtain $\delta^{\bf k=0}_{_M}=0$ and other sub-horizon modes $\delta^{\bf k\not=0}_{_M}$ in Eq.~(\ref{kdp+0}) are $e^{-Ht}$ damping solutions in time.  
Thus $\delta^{\bf k}_{_M}(t)=0$ for all modes $\bf k$ ($\omega_{\delta,\Delta}({\bf k})>H$) inside the horizon. For the Fourier transformation (\ref{fm1}) with a finite spatial volume $V$ bound by the horizon, all modes $\delta^{\bf k}_{_M}$ are expressed based on a complete set of states $|\bf{k}\rangle, \langle x|\bf{k}\rangle\propto e^{i{\bf k}{\bf x}}
$ and different modes $\bf{k}$ 
and $\bf{k'}$ 
are orthogonal $\langle\bf{k}|\bf{k'}\rangle=\delta_{\bf{k},\bf{k'}}$.
Therefore, the momentum mode solution 
$\delta^{\bf k}_{_M}(t)=0$ leads to spatial distribution 
$\delta_{_M}(\bf{x},t)=0$ inside the horizon.}

\section
{\bf Particle-antiparticle oscillation and horizon crossing}\label{mode0}

To study how horizon-crossing modes produce particle-antiparticle asymmetry, we first recall the usual approach to study the curvature (metric) perturbations $\delta\phi$ and their horizon crossings, which cause density perturbations for large-scale structure and galaxy formation, for example, \cite{Kolb1990}. Microscopic physics causally operates only on distance scales less than the Hubble radius, as the Hubble radius represents the distance a light signal can travel in an expansion time $\tau_{_H}=H^{-1}$. As a microscopic mode $\delta\phi_k$ of physical wavelength $(k/a)^{-1}> H^{-1}$, it crosses outside the horizon, decouples from microphysics and freezes in as a classical field $\delta\phi_k\approx {\rm constant}$. It is the phenomenon of microscopic mode $\delta\phi_k$ super-horizon crossing. When its physical wavelength $(k/a)^{-1}< H^{-1}$, the frozen mode reenters the horizon and returns to the fluctuating field coupling to microphysics. Such sub-horizon crossing phenomenon causes density perturbations for large-scale structure and galaxy formation. 
The perturbation mode super- or sub-horizon crossing is a dynamical process competition between the microscopic physics scale and the macroscopic horizon $H$ scale varying in time. 

Instead of the curvature perturbation modes $\delta\phi_k$, we study 
the similar dynamics and phenomenon for the density contrast perturbation modes $\delta^{\bf k}_{_M}$ (\ref{kdp+}) and $\Delta^{\bf k}_{_M}$ (\ref{kdp-}) whose microscopic physics scales are frequencies (\ref{fdp+},\ref{fdp-}). The $\delta^{0}_{_M}\not =0$ ($\delta^{0}_{_M}=0$) inside the horizon implies massive particle $X$ and antiparticle $\bar X$ asymmetry (symmetry). We will show their super- and sub-horizon crossings phenomena possibly accounting for 
dark-matter particle and antiparticle asymmetry and baryogenesis.

\subsection
{Under- and over-damped oscillating modes}\label{mode1}

We start with the analysis for the lowest-lying modes $\delta^0_{_M}$ and $\Delta^0_{_M}$ of 
Eqs.~(\ref{kdp+0}-\ref{fdp-0}). In general, we expect that for large frequencies 
$\omega_{\delta,\Delta} \gg (\Gamma_M+ 2H)$, 
the modes $\delta^0_{_M}$ and $\Delta^0_{_M}$ are underdamped oscillating inside the horizon.
While small frequencies $\omega_{\delta,\Delta} \ll (\Gamma_M+ 2H)$, the friction term 
$(\Gamma_M+ 2H)\dot\delta^0_{_M}$ dominates, 
the amplitudes of the modes $\delta^0_{_M}$ and $\Delta^0_{_M}$ 
are overdamped and frozen to be constants outside the horizon $\delta^0_{_M}\rightarrow {\rm const}$ and $\Delta^0_{_M}\rightarrow {\rm const}$, decoupling from microscopic physics. Horizon crossing occurs at $\omega_{\delta,\Delta} \approx (\Gamma_M+ 2H)$.
Following the usual approach for curvature perturbations,
we present a quantitative analysis to show the horizon-crossing phenomenon 
by using the simplified 
Eqs.~(\ref{kdp+0},\ref{kdp-0}) and (\ref{fdp-0}) 
in one dimension and assuming isotropic 
three-dimension oscillations. 


In terms of dimensionless variables 
$t\rightarrow \hat mt$, $\Gamma_M\rightarrow \Gamma_M/\hat m$ and 
$H\rightarrow H/\hat m$, we rewrite Eq.~(\ref{kdp+0}) in the form of the usual equation for a dampened harmonic oscillator,
\begin{eqnarray}
\ddot\delta^0_{_M} +2\zeta\omega_\delta \dot\delta^0_{_M} +\omega_\delta^2\delta^0_{_M}
&=& 0, 
\label{hosi}
\end{eqnarray} 
and the same for the mode $\Delta^0_{_M}$.
The underdamped frequency $\omega_\delta$ and the damping ratio $\zeta$ are,
\begin{eqnarray}
\omega^2_\delta  \equiv  2\Gamma_M H,\quad \zeta \equiv (\Gamma_M +2H)/(2\omega_\delta), 
\label{hosid}
\end{eqnarray}    
where $\Gamma_M=\chi \hat m \epsilon/(4\pi)$ (\ref{prate}) is almost a constant in time, see Fig.~\ref{AllBrR} for the $\mathcal M$-{\it episode} of the reheating epoch, and $H$-variation has the time scale $\tau_{_M}=H^{-1}$ from the Friedman equations (\ref{friedman}) in the reheating epoch. Thus, in the oscillating equation (\ref{hosid}), $\omega^2_\delta$ and $2\zeta\omega_\delta$ are slowly 
time-varying functions in the period 
of 
the time-oscillating modes $\Delta^0_{_M}$ and $\delta^0_{_M}$, the same discussions apply to 
high-energy modes $|{\bf k}|$. 

When $\Gamma_M=0$ and $\omega_\delta=0$, Eq.~(\ref{hosi}) yields a damping solution $\delta^0_{_M}\propto e^{-Ht}\rightarrow 0$ in time, and same discussions for $\Delta^0_{_M}$. There are no dynamics of $\delta^0_{_M}$ and $\Delta^0_{_M}$ modes' oscillating and horizon crossing. The new term $\Gamma_M\not=0$ determines modes' frequency and damping (\ref{hosid}) in comparison with the Hubble rate $H$ of horizon expansion. Thus, it plays a crucial role in the following discussions and results of producing particle and antiparticle asymmetry.


We further approximately treat the slowly time-varying $\omega^2_\delta$ and $2\omega_\delta\zeta$ as constants in time. 
In this circumstance, the approximate solution to Eq.~(\ref{hosi}) reads  
\begin{eqnarray}
\delta^0_{_M} \propto e^{-\omega_\delta\zeta t}e^{-i\omega_\delta(1-\zeta^2)^{1/2} t}.
\label{hosis}
\end{eqnarray}
We have the following physical situations:
\begin{enumerate}[(i)]
\item  in the underdamped case $(\zeta<1)$, i.e., $2\omega_\delta> \Gamma_M +2H$, 
the modes $\delta^0_{_M}$ and $\Delta^0_{_M}$ oscillate with smaller frequencies than 
$\omega_\delta$, their wavelengths are smaller than the horizon size, 
and amplitudes damped to zero inside the horizon. This 
underdamped case corresponds to the non-horizon-crossing case, i.e.,
the sub-horizon case (a) below Eq.~(\ref{Ndp+}). 
No particle-antiparticle asymmetry is produced inside the horizon. The damping time scale $(\omega_\delta\zeta)^{-1}=(\Gamma_M+H)^{-1}\approx \Gamma_M^{-1}$. 
\item in the overdamped case $(\zeta>1)$, i.e., $2\omega_\delta <\Gamma_M +2H$, the
$\delta^0_{_M}$ and $\Delta^0_{_M}$ modes' wavelengths are larger than the horizon size, and their amplitudes exponentially decay and return to steady states without oscillating. This 
overdamped case corresponds to the horizon-crossing case, i.e., the super-horizon case (b) below Eq.~(\ref{Ndp+}).  In the case $(\zeta \gg 1)$, the solution (\ref{hosis}) yields 
$\delta^0_{_M}, \Delta^0_{_M} \propto {\rm const.}$, indicating the amplitudes of 
modes $\delta^0_{_M}$ and $\Delta^0_{_M}$ are ``frozen'' to constants outside 
the horizon. It implies that the particle-antiparticle asymmetry could be produced inside the Horizon, as discussed below.
\end{enumerate}

\subsection{Lowest-lying oscillation mode crossing horizon}

The separatrix between two situations (i) and (ii) 
is defined at $\zeta =1$. At this separatrix,
the frequency $\omega_\delta$ and damping ratio 
$\zeta$ (\ref{hosid}) lead to 
\begin{eqnarray}
\Gamma_M=2H,
\label{ccon}
\end{eqnarray}
and the critical ratio of horizon radius and pair oscillating length 
\begin{eqnarray}
\frac{H^{-1}}{\omega^{-1}_\delta} = \left(\frac{2\Gamma_M}{H}\right)^{1/2}=2.
\label{hcri}
\end{eqnarray}
Such a critical ratio represents the horizon crossing of the zero 
modes $\delta^0_{_M}$ and $\Delta^0_{_M}$:
\begin{enumerate}[(a)]
\item  subhorizon  $\delta^0_{_M}$ and $\Delta^0_{_M}$ modes are {\it inside} the horizon for
$
(H^{-1}/\omega^{-1}_\delta) >2;
$
\item superhorizon  $\delta^0_{_M}$ and $\Delta^0_{_M}$ modes are {\it outside} the horizon for 
$
(H^{-1}/\omega^{-1}_\delta) <2.
$
\end{enumerate}
These results have clear physical meanings. Whether the oscillating modes 
are the subhorizon size or superhorizon size crucially depends on the 
``time-competition'' (\ref{ccon}) between the density perturbations $(\rho^{\pm}_{_M}\Leftrightarrow \rho^0_{_M})$ rate $\Gamma_M$ and the Hubble rate $H$ of spacetime expansion:
$\Gamma_M>H$ the modes stay inside the horizon; $\Gamma_M<H$ the modes stay outside 
the horizon. In other words, the modes stay inside (outside) 
the horizon, if they oscillate faster (slower) than the spacetime expanding 
rate, since they have (no) enough time to keep themselves inside the horizon.  
Consistently, the ``space-competition'' ratio (\ref{hcri}) 
of horizon size and mode wavelength shows (a) subhorizon-sized modes and (b) 
superhorizon-sized modes. 

The horizon crossing 
condition (\ref{hcri}) clearly depends on the functions 
$H(t)$ and $\Gamma_M(t)$ in different epochs of the Universe 
evolution. 
At the horizon crossing (\ref{hcri}), the rate  (\ref{prate}) is   
$\Gamma^{\rm cr}_M=\chi \hat m \epsilon_{\rm cr}/(4\pi)$, 
we find the horizon crossing condition $H_{\rm cr} = \Gamma^{\rm cr}_M/2$ (\ref{ccon}) for the zero mode  
\begin{eqnarray}
(H_{\rm cr}/\chi\hat m)= (1/8\pi)\epsilon_{\rm cr},
\label{ccons}
\end{eqnarray}
where $H_{\rm cr}$ and $\epsilon_{\rm cr}$ stand for the Hubble rate and $\epsilon$-rate 
at the horizon crossing.
The LHS of Eq.~(\ref{ccons}) is the ratio of massive pair plasma layer width and horizon radius, e.g., $(\lambda_m/H_{\rm cr}^{-1})$. While the RHS of Eq.~(\ref{ccons}) is the 
$\epsilon$-rate (\ref{dde0}) depending on the horizon evolution. They are equal when the horizon crossing occurs.

\comment{We end this section by noting that the term 
$\dot \Gamma_M\propto \dot \epsilon >0$ in Eqs.~(\ref{dp+},\ref{dp-}) 
might not be negligible for fast time-increasing $\epsilon$-rate in the ${\mathcal P}$-episode of the reheating. The effective oscillating frequencies $\omega_\delta=2H\Gamma_M + \dot \Gamma_M> 2H\Gamma_M$ (wavelengths $\omega^{-1}_\delta$) become 
larger (smaller). Therefore the critical ratio (\ref{hcri}) should be larger than $2$, compared with the case $\dot \Gamma_M=0$.}

\section
{Oscillating amplitudes at horizon crossing}\label{ampath}

To quantify symmetric and asymmetric oscillating amplitudes at horizon crossing, we define
the root-mean-square ($rms$) density fluctuations by 
\begin{eqnarray}
\bar\delta_{_M}\equiv \langle\delta_{_M}({\bf x})\delta^\dagger_{_M}({\bf x})\rangle^{1/2},\quad
\bar\Delta_{_M}\equiv \langle\Delta_{_M}({\bf x})\Delta^\dagger_{_M}({\bf x})\rangle^{1/2}
\label{rms}
\end{eqnarray}
where $\langle\cdot\cdot\cdot\rangle = V^{-1}\int d^3x(\cdot\cdot\cdot)$ indicates the average all states over the space.
The use of Fourier transformations (\ref{fm1}) and (\ref{fm2}) yields
\begin{eqnarray}
\bar\delta^{~2}_{_M} =
\frac{1}{V}\sum_{{\bf k},{\bf k}'} 
\delta^{\bf k}_{_M}\delta^{{\bf k}'\dagger}_{_M}\delta_{{\bf k},{\bf k}'}=\frac{1}{V}\sum_{\bf k} 
|\delta^{\bf k}_{_M}|^2
\label{rms1}
\end{eqnarray}
and the same for $\bar\Delta^{~2}_{_M}$, 
where the dimensionless $\delta_{{\bf k},{\bf k}'}=V^{-1}\int d^3x e^{i{\bf x}({\bf k}-{\bf k}')}$ 
is the Kronecker delta function of discrete variables ${\bf k}$ and ${\bf k}'$.  

\subsection{Lowest-lying mode amplitudes at horizon crossing}

Considering contribution only from the lowest-lying modes (ground state) of underdamped oscillating 
$\delta^0_{_M}$ and $\Delta^0_{_M}$, we approximately obtain from Eq.~(\ref{rms1})    
\begin{eqnarray}
\bar\delta^{~2}_{_M}\approx V^{-1}|\delta^0_{_M}|^2,\quad \bar\Delta^{~2}_{_M}\approx V^{-1}|\Delta^0_{_M}|^2.
\label{rms2}
\end{eqnarray}
At a fixed time $t$, the amplitudes 
$|\delta^0_{_M}|^2$ and $|\Delta^0_{_M}|^2$ of the 
lowest-lying modes of the underdamped harmonic oscillator (\ref{hosi}) can be expressed by the oscillation
characteristic length scale $1/(2\hat m\omega_{\delta,\Delta})^{1/2}$,  
\begin{eqnarray}
|\delta^0_{_M}|^2\approx 1/(2\hat m\omega_\delta)^{3/2},\quad |\Delta^0_{_M}|^2\approx 1/(2\hat m\omega_\Delta)^{3/2},
\label{rmss}
\end{eqnarray}
see, e.g., 
Refs.~\cite{Sakurai2020}, \cite{landaunon} and \cite{Hollands2002}. If the wavelength of the lowest-lying mode is much smaller than the 
horizon radius, the mode will evolve adiabatically, and Eq.~(\ref{rmss}) will continue to hold at later times. 
From Eq.~(\ref{rms2}), we obtain the root-mean-square of density fluctuations
\begin{eqnarray}
\bar\delta^{~2}_{_M}=\bar\Delta^2_{_M} \approx  \frac{1}{4\pi(2\hat m)^{3/2}} \frac{3 H^3}{(2H\Gamma_M)^{3/4}}.
\label{finr}
\end{eqnarray}
At the other extreme, if the wavelength of the lowest-lying mode is much larger than the horizon radius, the oscillator will overdamp, and the oscillating amplitudes $\bar \delta_{_M}$ and $\bar \Delta_{_M}$ will remain constants in time.
These constants at the horizon crossing $H_{\rm cr}=\Gamma^{\rm cr}_M/2$ (\ref{ccons}) are
\begin{eqnarray}
\bar\delta_{_M}=\bar\Delta_{_M} \approx \left(\frac{3}{32\pi}\right)^{1/2}\left(\frac{H_{\rm cr}}{\hat m}\right) ^{3/4}
= \left(\frac{3}{32\pi}\right)^{1/2}\left(\frac{\chi\epsilon_{\rm cr}}{8\pi }\right) ^{3/4},
\label{finr1}
\end{eqnarray}
whose values depend on the Hubble rate $H$ or the $\epsilon$-rate values at the 
horizon crossing. 

\subsection{Particle-antiparticle asymmetry due to horizon crossing}

As a result, using Eqs.~(\ref{delta1},\ref{delta2}), we explicitly write the result (\ref{finr1}) as, 
\begin{eqnarray}
\rho^{+}_{_M}-\rho^{-}_{_M}&= & \bar\delta_{_M}\rho^H_{_M},\label{finr2a}\\
\rho_{_M}-\rho^H_{_M}&= & \bar\Delta_{_M}\rho^H_{_M},
\label{finr2b}
\end{eqnarray}
where the right-handed sides are in the sense of $rms$. In other words,  
$\bar\delta_{_M}\rho^H_{_M}$ represents the spatial fluctuations in the number of particles or antiparticles (compositions) 
per comoving volume, and $\bar\Delta_{_M}\rho^H_{_M}$
represents the spatial fluctuations in the number of pairs per comoving volume. 
This result physically implies the following two consequences due to the 
particle-antiparticle oscillations 
at the horizon crossing: 
\begin{enumerate}[(i)]
\item 
In the case that the $\delta^0_{_M}$ is an underdamped oscillating mode inside 
the horizon, its root-mean-squared ($rms$) value  $\bar\delta_{_M}$ vanishes, 
indicating all particles and antiparticles are inside the horizon, no net particle number appears with respect to a subhorizon observer, i.e., an observer inside the horizon. This underdamped case corresponds to the non-horizon-crossing case (i) below Eq.~(\ref{hosis}) and the sub-horizon case (a) below Eq.~(\ref{Ndp+}).
There is no particle-antiparticle asymmetry inside the horizon. 
\item 
In the case that the $\delta^0_{_M}$ is an overdamped oscillating mode frozen outside 
the horizon, its root-mean-squared ($rms$) value  $\bar\delta_{_M}$ does not vanish. It  
indicates that some particles (or antiparticles) are outside the horizon.
Thus net particle number appears with respect to the subhorizon observer. This overdamped case corresponds to the 
horizon-crossing case (ii) below Eq.~(\ref{hosis}) and the 
super-horizon case (b) below Eq.~(\ref{Ndp+}).
There is a particle-antiparticle asymmetry inside the horizon.
Note that the ``frozen'' amplitudes of $\bar \delta_{_M}$ and $\bar \Delta_{_M}$ (\ref{finr1}) are very small for $\hat m\gg H_{\rm cr}$.
\end{enumerate}
It is important to note that the summed number of particles and antiparticles inside and 
outside the horizon is zero in both cases, as required by total particle number conservation. In the second case (ii), the positive (negative) net number of particles and antiparticles viewed by the subhorizon observer is
equal to the negative (positive) net number of particles and antiparticles 
outside the horizon. The asymmetric perturbation 
$\bar\delta_{_M}$ occurring at the horizon crossing describes particle and antiparticle number asymmetry inside the horizon.

The lowest-lying mode $\delta^0_{M}$ inside the horizon represents an under-damping oscillation between particles and antiparticles and its dampened amplitude vanishes within the horizon. As a result, the averaged net number of particles is zero at the time 
period $t\sim H^{-1}$, and particle and antiparticle symmetry is preserved. 
Instead, the lowest-lying mode $\delta_{M}$ outside the horizon means
that its amplitude is frozen to a constant $\delta^0_{M}={\rm const.}\not= 0$. 
This implies that the subhorizon observer should observe a non-vanishing net particle number associating the horizon surface, representing the modes of constant amplitude outside the horizon. Such mode horizon crossing indicates the occurrence of particle and antiparticle asymmetry. The same discussions apply to the pair-density perturbation mode $\Delta^0_{M}$ (\ref{delta1}), which however does not violate the symmetry of particle and antiparticle.
In the next section, we must examine when such mode horizon crossing occurs in the reheating epoch.

\comment{Suppose that the power spectrum $|\delta^{\bf k}_{_M}|^2$ is isotropic, i.e., depends only upon $|{\bf k}|$, this is consistent with the simplified 
analysis of one-dimension harmonic oscillator for three-dimension harmonic 
oscillator.
\begin{eqnarray}
\bar\delta^{~2}_{_M}=\sum_{{\bf k}_\perp} \sum_{|{\bf k}|}
|\delta^{|{\bf k}|}_{_M}|^2=\frac{4\pi H^{-2}}{(2\pi)^3}  \sum_{|{\bf k}|}
|\delta^{|{\bf k}|}_{_M}|^2\approx \frac{4\pi H^{-2}}{(2\pi)^3}
|\delta_{_M}|^2 
\label{rms2}
\end{eqnarray}
where the number of transverse states $\sum_{{\bf k}_\perp}=4\pi H^{-2}/(2\pi)^2$ and we consider the lowest lying mode $|{\bf k}|=0$. $|\delta_{_M}|^2=1/(2\omega_\delta)$
}

\section
{\bf Subhorizon crossing in pre-reheating}\label{hcross0sub}

To identify the subhorizon crossing at preheating ${\mathcal P}$-episode, when the reheating starts, 
we examine that (i) the modes $\delta^0_{M}$ and $\Delta^0_{M}$ are superhorizon size 
in the pre-inflation and inflation epochs before the  ${\mathcal P}$-episode; (ii) they become subhorizon size in the massive pair 
oscillating ${\mathcal M}$-episode, after the ${\mathcal P}$-episode, see Fig.~\ref{AllBrR}.

\subsection{Particle-antiparticle asymmetry in pre-inflation and inflation}

\begin{figure*}[t]
\includegraphics[height=5.5cm,width=7.4cm]{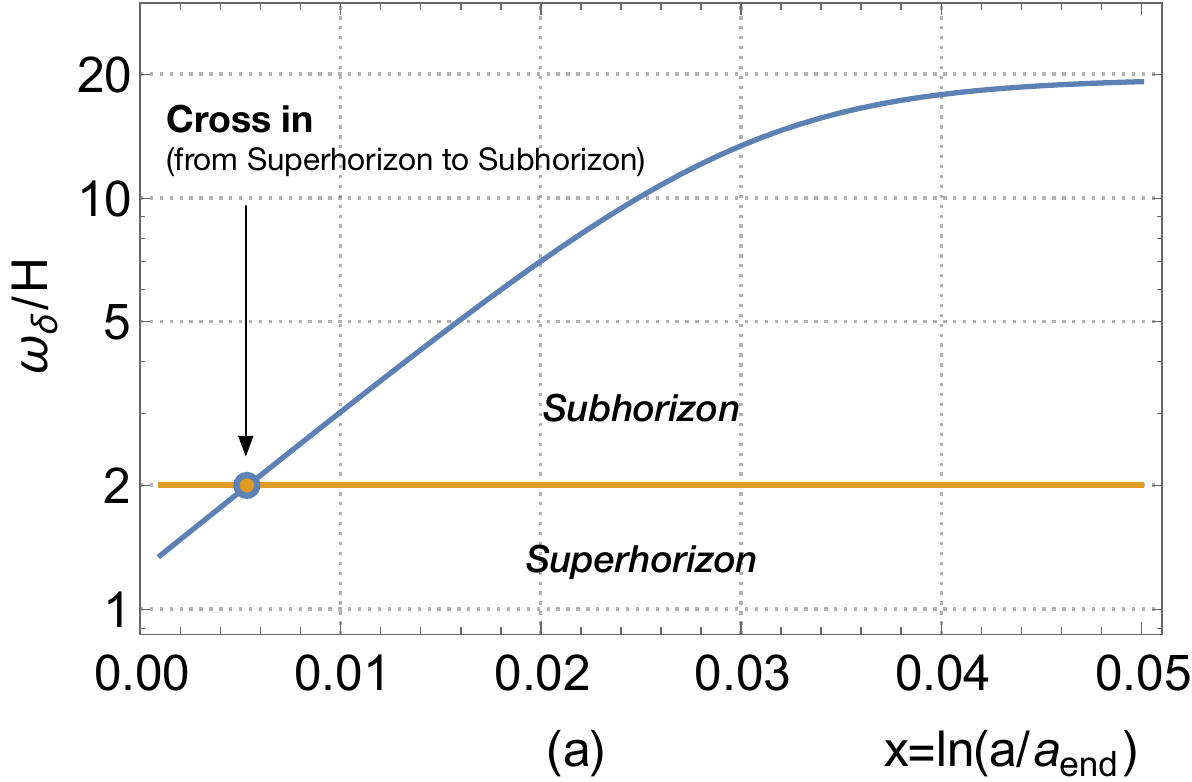}\hspace{0.333cm}
\includegraphics[height=5.5cm,width=7.4cm]{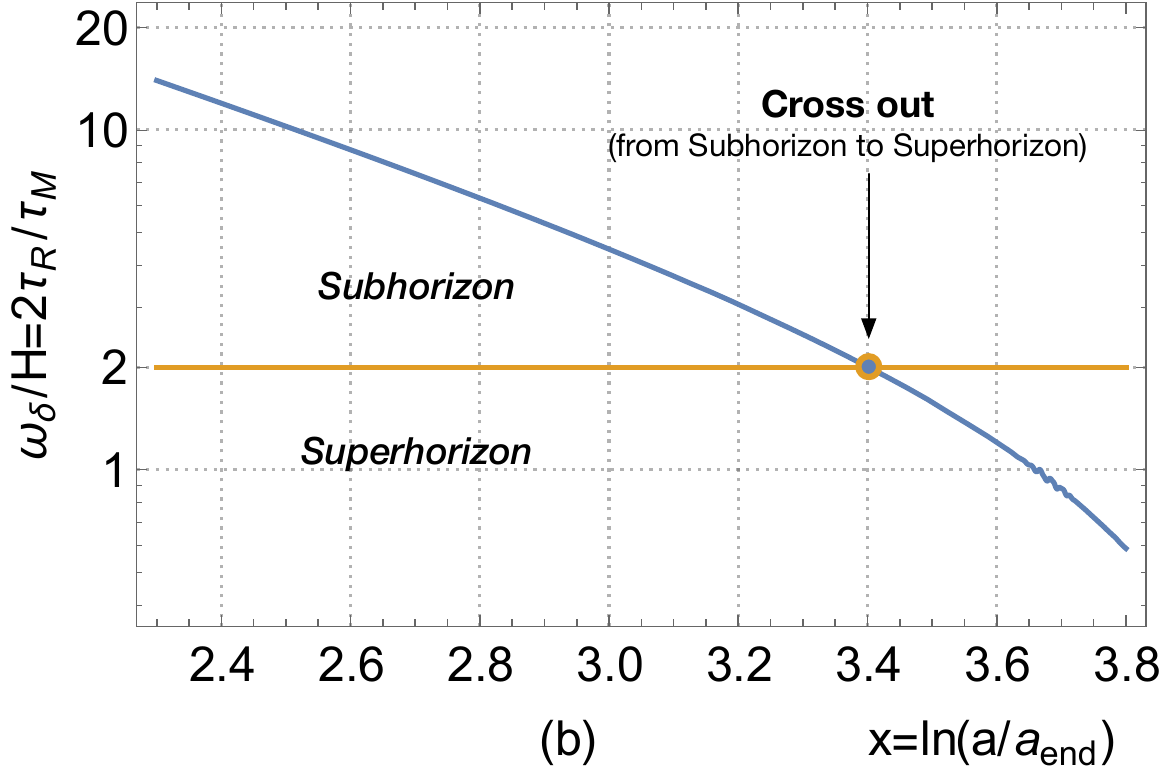}
\caption{ (Color Online). In these two figures (a) and (b), we use the previous numerical results 
$H$ and $\Gamma_M$ obtained by solving Eqs.~(6.6-6.11) with the parameters $\hat m/m_{\rm pl}=27.7$ and $g_{_Y}^2=10^{-9}$ in Ref.~\cite{Xue2023a}. 
In terms of the $e$-folding number $x=\ln(a/a_{\rm end})$,
the ratio $\omega_\delta/H$ 
(\ref{ifinside}) of the horizon size $H^{-1}$ and the oscillating length 
$\omega^{-1}_\delta$ is plotted (blue),
compared with the horizon crossing $\omega_\delta/H=2$ (orange).
The superhorizon (subhorizon) is below (above) the orange horizontal line. Left (a): The ratio $\omega_\delta/H=(2\Gamma_M/H)^{1/2}$ (\ref{hcri}) (blue) 
in the preheating ${\mathcal P}$-episode. The modes (blue line) evolve 
from superhorizon to subhorizon and cross at $\ln(a_{\rm crin}/a_{\rm end})\approx 0.005$ and $a_{\rm crin}\approx 1.01 ~a_{\rm end}$. 
Right (b): The ratio $\omega_\delta/H\approx (2\tau_{_R}/\tau_{_M})$ (\ref{routside}) (blue) 
in the genuine reheating ${\mathcal R}$-episode. 
The modes (blue line) evolve 
from subhorizon to superhorizon and cross at $\ln(a_{\rm crout}/a_{\rm end})\approx 3.4$ and $a_{\rm crout}\approx 30 a_{\rm end}$. Whereas, the reheating occurs at $\ln(a_{_R}/a_{\rm end})\approx 3.8$ 
and $a_{_R}\approx 45~a_{\rm end}$ from Fig.~7 (a) of Ref.~\cite{Xue2023a}.
Thus, $a_{\rm crout}\lesssim a_{_R}$. It shows the crossing occurs before the massive pair domination (${\mathcal M}$-episode) ends. 
This is the reason that we neglect the decay rate $\Gamma^{\rm de} \ll \Gamma_M$ in analyzing Eq.~(\ref{continuum}). Also we neglect the term $\dot \Gamma_M\propto \dot\epsilon \approx 0$ in acoustic equations (\ref{dp+},\ref{dp-}), since the subhorizon crossing is near inflation end and the superhorizon crossing is near ${\mathcal M}$-episode end where $\dot \epsilon \approx 0$.}\label{crossingf}
\end{figure*}

In the pre-inflation and inflation epoch $H >\Gamma_M$ and $\epsilon \ll 1$, when $x=\ln(a/a_{\rm end}) \leq 10^{-3}$ in Fig.~\ref{AllBrR}, the modes $\delta^0_{_M}$ and 
$\Delta^0_{_M}$ are outside the horizon, corresponding to the overdamped case 
$(\zeta >1)$. This can also be seen by
the ratio of horizon radius $H^{-1}$ and pair oscillating length $\omega_\delta^{-1}$ (\ref{fdp-0}),
\begin{eqnarray}
\frac{H^{-1}}{\omega^{-1}_\delta} = \left(\frac{H^{-2}}{2^{-1}\Gamma_M^{-1}H^{-1}}\right)^{1/2}< \left(\frac{\Gamma_M^{-1}}{2^{-1}\Gamma_M^{-1}}\right)^{1/2}=2^{1/2}, 
\label{ifoutside}
\end{eqnarray}
indicating $H^{-1}/\omega^{-1}_\delta <2$. We use the numerical results in Ref.~\cite{Xue2023a} to show the ratio (\ref{ifoutside}) in Fig.~\ref{crossingf} (a), 
where the blue line $H^{-1}/\omega^{-1}_\delta$ is below the orange line $2$. 
This implies that in the pre-inflation and inflation, 
the modes $\delta^0_{_M}$ and $\Delta^0_{_M}$ wavelengths are larger than the horizon size, 
their amplitudes exponentially decay and return to steady states without oscillating. It 
means that the modes $\delta^0_{_M}$ and $\Delta^0_{_M}$ are superhorizon size,
and their amplitudes are ``frozen'' 
to constants outside the horizon. 
With respect to the subhorizon observer, it means that   
$\bar\delta_{_M}$ (\ref{finr2a}) does not vanish 
and the particle-antiparticle symmetry breaks in the pre-inflation 
and inflation epoch ($a<a_{\rm end}$).
It is the case corresponding to the super-horizon case (b) below Eq.~(\ref{Ndp+}).

\subsection{Particle-antiparticle symmetry in massive pair episode}

In the ${\mathcal M}$-episode $H<\Gamma_M/2$ and $\epsilon\approx 3/2$, when $7\cdot 10^{-2} \lesssim \ln(a/a_{\rm end}) \lesssim 3.6$ of Figs.~\ref{AllBrR}, \ref{crossingf} (a) and (b), 
we find that the 
modes $\delta^0_{_M}$ and $\Delta^0_{_M}$ are inside the horizon. This is because the 
the ratio of horizon radius and pair oscillating length is larger than $2$,
\begin{eqnarray}
\frac{H^{-1}}{\omega^{-1}_\delta} = \left(\frac{H^{-2}}{2^{-1}H^{-1}\Gamma_M^{-1}}\right)^{1/2}> \left(\frac{2\Gamma_M^{-1}}{2^{-1}\Gamma_M^{-1}}\right)^{1/2}=2.
\label{ifinside}
\end{eqnarray}
It means that the modes $\delta^0_{_M}$ and $\Delta^0_{_M}$ are subhorizon 
sized and underdamped oscillations, corresponding to the underdamped case $(\zeta <1)$. 
We use the numerical results in Ref.~\cite{Xue2023a} to show the ratio (\ref{ifinside}) in Fig.~\ref{crossingf} (a), 
where the blue line $H^{-1}/\omega^{-1}_\delta$ is above the orange line $2$.
Therefore, in the ${\mathcal M}$-episode the particle-antiparticle asymmetric 
mode $\delta^0_{_M}$ and symmetric mode $\Delta^0_{M}$ are well inside the horizon. They 
behave as underdamped oscillating waves whose 
dampened amplitudes vanish within the horizon $H^{-1}$. 
Their root-mean-square density fluctuations (\ref{rms}) 
$\bar \delta_{_M}=0$ and $\bar \Delta_{_M}=0$ vanish.
It means that with respect to the subhorizon observer, the asymmetric 
perturbation (\ref{finr2a}) vanishes 
and the particle-antiparticle symmetry holds in the ${\mathcal M}$-episode.

The  
$\bar \delta_{_M}=0$ and $\bar \Delta_{_M}=0$ equivalently correspond to 
the averaged $\langle \delta^0_{_M}\rangle=0$ and 
$\langle \Delta^0_{_M}\rangle=0$ over the Hubble time scale $\tau_{_H}\sim H^{-1}$, which is larger than the perturbation time scale $\tau_{_M}$ (\ref{prate}). The $\langle \Delta^0_{M}\rangle=0$ corresponds to the detailed balance of perturbations 
$\rho^{\pm}_{_M}\Leftrightarrow \rho^H_{_M}/2$ 
in the rate equation (\ref{continuum}) inside the horizon. Therefore, the net number of particles and antiparticles is zero and 
the particle-antiparticle symmetry is preserved in the ${\mathcal M}$-episode.

\subsection{Subhorizon crossing point in pre-reheating episode}

Form the superhorizon size (\ref{ifoutside}) in the 
pre-inflation and inflation epoch to the 
subhorizon size (\ref{ifinside}) in the ${\mathcal M}$-episode, 
the modes $\delta_{_M}$ and $\Delta_{_M}$ cross at least once the horizon.
Because the $H$ and $\Gamma_M$ vary 
monotonically, Equations (\ref{ifoutside}) and (\ref{ifinside}) 
show that one horizon crossing point $\omega_\delta=H$ locates at $H=\Gamma_M/2$ 
in the pre-reheating ${\mathcal P}$-episode, when $10^{-3}\lesssim \ln(a/a_{\rm end}) \lesssim 7\cdot 10^{-2}$ of Figs.~\ref{AllBrR}, \ref{crossingf} (a) and (b). 
Using Eq.~(\ref{ccon}), we find the subhorizon crossing scale $H_{\rm crin}$ and oscillating 
frequency $\omega^{\rm crin}_{\delta}$ at the horizon-crossing scale factor $a_{\rm crin}$,
\begin{eqnarray}
H_{\rm crin}=\Gamma_M/2\approx H_{\rm end}/2,\quad \omega^{\rm crin}_{\delta}=(2\Gamma_M H)^{1/2}\approx H_{\rm end}, 
\label{cconf}
\end{eqnarray}
which are the same order of the inflation end scale $H_{\rm end}\approx \Gamma_M$. 
In Fig.~\ref{crossingf} (a), using previous numerical results, we plot the ratio 
$H^{-1}/\omega^{-1}_\delta=(2\Gamma_M/H)^{1/2}$ (\ref{ifinside}), 
starting from the inflation end $H_{\rm end}$, to show the subhorizon crossing occurs 
at $x=\ln(a_{\rm crin}/a_{\rm end})\approx 5\times 10^{-3}$, i.e., 
$a_{\rm crin}\approx 1.01 a_{\rm end}$ in the preheating ${\mathcal P}$-episode. 
It shows that the subhorizon crossing occurs right after the inflation end, $H_{\rm crin}> H_{\rm end}$ and 
$a_{\rm crin}\gtrsim a_{\rm end}$.

In the pre-inflation and inflation epoch, the subhorizon observer views the 
particle-antiparticle asymmetry because some of the particles or antiparticles are outside 
the horizon. These superhorizon particle or antiparticle modes cross back to the horizon in the ${\mathcal P}$-episode. The subhorizon observer views the particle-antiparticle symmetry in the ${\mathcal M}$-episode because all particles and antiparticles are inside the horizon.  
At such subhorizon crossing $a_{\rm crin}$, the number of particles or antiparticles
can be calculated as follows. The numerical value 
$\epsilon_{\rm cr}=\epsilon_{\rm crin}\approx 1.0\times 10^{-2}$ at the subhorizon crossing $a_{\rm crin}$  
can be found from Fig.~\ref{AllBrR} and \ref{crossingf} (a). 
We use Eq.~(\ref{finr1}) to 
calculate the asymmetric and symmetric pair density perturbations
\begin{eqnarray}
\bar\delta^{\rm crin}_{_M}=\bar\Delta^{\rm crin}_{_M}\approx 4.33\times 10^{-6}, 
\label{deltain}
\end{eqnarray}
at the subhorizon crossing $a_{\rm crin}$.
Then 
the net particle density perturbation (\ref{finr2a}) and the pair 
density perturbation (\ref{finr2b}) at the subhorizon crossing (\ref{cconf}) are given by,
\begin{eqnarray}
\rho^{+}_{_M}-\rho^{-}_{_M}&=& \bar\delta^{\rm crin}_{_M}\rho^H_{_M}\approx \bar\delta^{\rm crin}_{_M}(\hat m^2H^2_{\rm end})/2,\label{finr2a1}\\
\rho_{_M}-\rho^H_{_M}&= & \bar\Delta^{\rm crin}_{_M}\rho^H_{_M}\approx \bar\Delta^{\rm crin}_{_M}(\hat m^2H^2_{\rm end})/2.
\label{finr2b1}
\end{eqnarray}
They are about $10^{-3}$ in unit of the characteristic density 
$\rho^c_{\rm end}=3m_{\rm pl}H^2_{\rm end}$ 
at inflation end. 
The particle-antiparticle asymmetry (\ref{finr2a1}) is on the horizon surface 
at the subhorizon crossing. It represents the superhorizon particle-antiparticle
asymmetry in the pre-inflation and inflation epochs. 
It also represents the subhorizon restoration of the practice-antiparticle 
symmetry in the ${\mathcal M}$-episode. This is because the total number of particles and antiparticles inside and outside the horizon is zero. 

As shown in Fig.~\ref{crossingf} (a), after the subhorizon crossing (\ref{cconf}), the subhorizon sized modes $\delta^0_{_M}$ and 
$\Delta^0_{_M}$ remain inside the horizon as underdamped oscillating modes 
in the ${\mathcal M}$-episode and preserve particle and antiparticle symmetry, until
they undergo the superhorizon crossing in the  genuine reheating ${\mathcal R}$-episode. 

\section{Superhorizon crossing in genuine reheating}\label{lcross}
 
We will show the modes $\delta^0_{_M}$ and $\Delta^0_{_M}$ superhorizon crossing
in the transition period from the massive pair domination ${\mathcal M}$-episode to 
the genuine reheating ${\mathcal R}$-episode when unstable massive pairs predominately 
decay into relativistic particles (w.r.t the reheating temperature) in the SM and dark matter particle contents. Such a superhorizon crossing produces particle-antiparticle asymmetry, giving rise to an initial condition for the standard cosmology after the reheating.

\subsection{Superhorizon crossing point in massive pair episode}
 
In the genuine reheating ${\mathcal R}$-episode, the horizon scale $H$ is mainly 
determined by unstable pairs' decay since the pair decay rate 
$\Gamma_M^{^{\rm de}}$ (time $\tau{_{_R}}$) is much larger (shorter) than the rate $\Gamma_M$ (time $\tau{_{_M}}$) of massive
pair perturbations. 
In Eq.~(\ref{continuum}), the decay term $\Gamma_M^{^{\rm de}}\rho^{\pm}_{_M}$ is dominant, 
and the density perturbations $\rho^{\pm}_{_M} \Leftrightarrow \rho^{0}_{_M}$ 
are no longer relevant. Unstable massive pairs rapidly decay to relativistic particles.

We adopt the horizon size $H^{-1}$ approximately determined by the reheating scale $H_{\rm RH}$ at the scale factor 
$a{_{_R}}$, 
\begin{eqnarray}
H^2_{\rm RH}=(2\tau_{_R})^{-2}=(\Gamma_M^{^{\rm de}}/2)^2,
\label{reheatingscale}
\end{eqnarray}
which is the first equation in (5.74) of Ref.~\cite{Kolb1990} and Eq.~(7.31) of Ref.~\cite{Xue2023a}. 
To examine the modes $\delta^0_{_M}$ and $\Delta^0_{_M}$ are 
subhorizon size or superhorizon size in the ${\mathcal R}$-episode, we use the criteria (\ref{ccon}) or (\ref{hcri}),
\begin{eqnarray}
\frac{H^{-1}}{\omega^{-1}_\delta} = \left(\frac{2\Gamma_M}{H}\right)^{1/2}\approx 2 (\tau_{_R}\Gamma_M)^{1/2}=2\left(\frac{\Gamma_M}{\Gamma_M^{^{\rm de}}}\right)^{1/2} <2.
\label{routside}
\end{eqnarray}
The last inequality or $\Gamma_M< 2H_{\rm RH} \approx \Gamma_M^{^{\rm de}}$
shows that the modes $\delta^0_{_M}$ and $\Delta^0_{_M}$ are superhorizon sizes 
in the ${\mathcal R}$-episode. 

From the subhorizon-sized modes in the ${\mathcal M}$-episode to the superhorizon-sized modes 
in the ${\mathcal R}$-episode, the $\delta^0_{_M}$ and 
$\Delta^0_{_M}$ modes' superhorizon crossing 
must occur in the transition period from the ${\mathcal M}$-episode to the ${\mathcal R}$-episode. The superhorizon crossing occurs at 
$H^{-1}_{\rm crout}/\omega^{-1}_\delta=2$, yielding
\begin{eqnarray}
 H_{\rm crout} = \Gamma_M^{^{\rm crout}}/2\approx \Gamma_M^{^{\rm de}}/2,\quad \Gamma^{^{\rm crout}}_M\approx \Gamma_M^{^{\rm de}},
\label{routside1}
\end{eqnarray}
consistently with the horizon crossing condition (\ref{ccon}) or (\ref{hcri}). 
The approximation $\Gamma^{^{\rm crout}}_M\approx \Gamma_M^{^{\rm de}}$ comes from 
$H_{\rm crout}\approx H_{\rm RH}=(\Gamma_M^{^{\rm de}}/2)$. It implies that 
the superhorizon crossing occurs at 
\begin{eqnarray}
H_{\rm crout}\gtrsim H_{\rm RH}
,\quad a_{\rm crout}\lesssim  a_{_R},
\label{rhc}
\end{eqnarray}
near to the 
reheating scale $H_{\rm RH}$ and 
$a_{_R}$. 

To verify such a superhorizon crossing point (\ref{routside1}), 
we use previous numerical results to plot the ratio $H^{-1}/\omega^{-1}_\delta\approx 2\tau_{_R}/\tau_{_M}$ of the horizon radius and the $\delta^0_{_M}$ mode oscillating length 
in Fig.~\ref{crossingf} (b). 
It shows that the ratio $H^{-1}/\omega^{-1}_\delta$ 
(blue line) varies from subhorizon ($>2$) to superhorizon ($<2$). 
The superhorizon crossing 
$H_{\rm crout}$ and $a_{\rm crout}$ (\ref{rhc}) are at the ${\mathcal M}$-episode end, which are indeed close to the genuine reheating $H_{\rm RH}$ and 
$a_{_R}$ (\ref{reheatingscale}).

\comment{The ratio (\ref{routside}) of horizon radius and pair wavelength 
is smaller than $2$,
\begin{eqnarray}
\frac{H^{-1}}{\omega^{-1}_\delta} \approx  \left(\frac{H^{-1}}{\tau_{_M}}\right)\approx  \left(\frac{2\tau_{_R}}{\tau_{_M}}\right)<2, \quad \tau_{_R}<\tau_{_M}
\label{routside}
\end{eqnarray}
for $\tau_{_R}<\tau_{_M}$, 
}

\subsection{Particle-antiparticle asymmetry occurs in reheating epoch}

As illustrated in Figs.~\ref{crossingf} (b), the modes $\delta^0_{M}$ and $\Delta^0_{M}$ 
have the superhorizon crossing (\ref{rhc}) in the genuine reheating
${\mathcal R}$-episode. The subhorizon observer views: 
\begin{enumerate}[(i)]
\item
the particle and antiparticle symmetry in the massive pair ${\mathcal M}$-episode; 
\item
superhorizon crossing (\ref{rhc}) produces the particle and antiparticle asymmetry in the genuine 
reheating ${\mathcal R}$-episode.
\end{enumerate}
Due to the superhorizon crossing at $a_{\rm crout}\lesssim a_{_R}$, the $\delta^0_{_M}$ is an overdamped oscillating mode 
frozen outside the horizon, its root-mean-squared ($rms$) value  $\bar\delta_{_M}$ does not 
vanish. It indicates that some particle (or anti-particle) modes are frozen outside the horizon.
From Eq.~(\ref{finr1}), we obtain the asymmetric and symmetric pair density 
perturbations at the superhorizon crossing (\ref{routside1})
\begin{eqnarray}
\bar\delta^{\rm crout}_{_M}=\bar\Delta^{\rm crout}_{_M}=2.31\times 10^{-4}, 
\label{deltaout}
\end{eqnarray}
by using $\epsilon_{\rm cr}=\epsilon_{\rm crout} \approx 2$ 
in the genuine reheating ${\mathcal R}$-episode.
Based on Eq.~(\ref{finr2a}), 
we calculate the particle-antiparticle
asymmetric number density, i.e., the net number density of particles and antiparticles, 
\begin{eqnarray}
\delta n^{\rm crout}_{_M}=\frac{\rho^{+}_{_M}-\rho^{-}_{_M}}{2\hat m} &= & \bar\delta^{\rm crout}_{_M}n^H_{_M}\big|_{\rm crout},\label{net0}
\end{eqnarray}
at the superhorizon 
crossing  $H_{\rm crout}\gtrsim H_{\rm RH}$ and $a_{\rm crout}\lesssim a_{_R}$.
As a result,
we approximately obtain the net number density of particles and antiparticles, 
\begin{eqnarray}
\delta n^{\rm crout}_{_M}=\frac{\rho^{+}_{_M}-\rho^{-}_{_M}}{2\hat m}\Big|_{\rm crout} &\approx & 2.31\times 10^{-4}n^H_{_M}\big|_{\rm crout},\label{net}
\end{eqnarray}
and $
n^H_{_M}\big|_{\rm crout}\approx \chi \hat m H^2_{\rm RH}$.
Analogously, we obtain  from Eq.~(\ref{finr2b}) the pair number 
density perturbation of massive particles,
\begin{eqnarray}
\Delta n^{\rm crout}_{_M}=\frac{\rho_{_M}-\rho^H_{_M}}{2\hat m}\Big|_{\rm crout} &\approx & 2.31\times 10^{-4}n^H_{_M}\big|_{\rm crout}.\label{net1}
\end{eqnarray}
The asymmetric modes $\delta^0_{M}$
carried the net particle number has been frozen in the superhorizon.  

It is important to note that the total number of particles and antiparticles inside and 
outside the horizon is zero and preserved. The positive (negative) net number of particles and antiparticles inside the horizon is
equal to the negative (positive) net number of particles and antiparticles 
outside the horizon. The latter is described by the asymmetric perturbation 
$\bar\delta_{_M}$ on the surface of the superhorizon crossing (\ref{deltaout}). 

\begin{figure}[t]
\centering
\includegraphics[height=7.0cm,width=10.8cm]{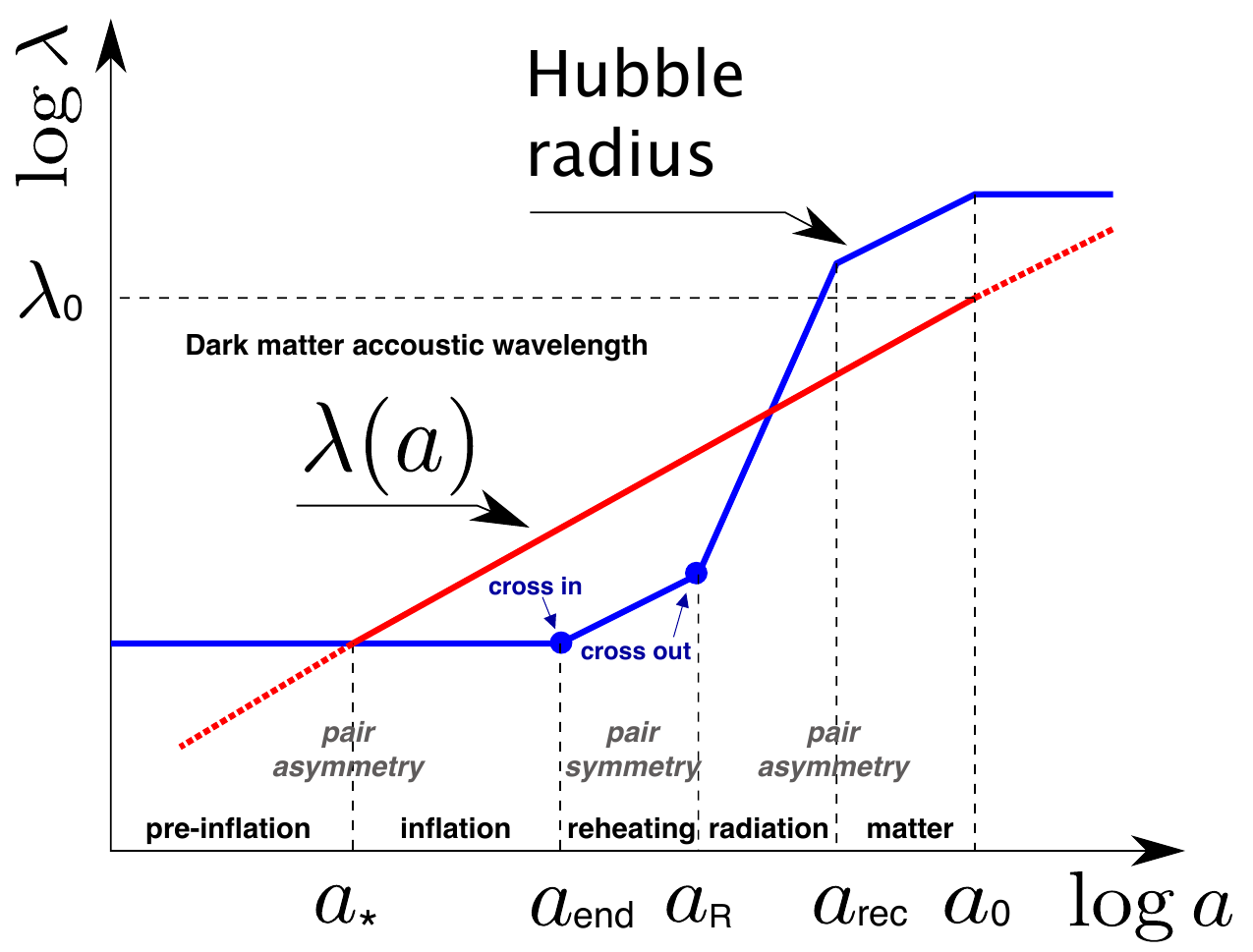}
\caption{We make this figure by modifying Fig.~1 in Ref.~\cite{Mielczarek2011} to illustrate 
schematic evolution of the Hubble radius $H^{-1}$ and the physical wavelength $\lambda(a)$ of dark matter acoustic wave. The wavelength 
$\lambda_0=\lambda(a_0)$ at the present time $a_0=1$ crossed the Hubble horizon $H_*$ at the early time $a_*$, 
fixed by the physically interesting length scale $\lambda_0=\lambda_*=k_*^{-1}$. The pre-inflation $a>a_*$, the inflation 
$a_*<a<a_{\rm end}$,
the reheating $a_{\rm end}<a<a_{_R}$,
and the recombination at $a_{\rm rec}$. The subhorizon ``cross in'' point corresponds to the 
point in Fig.~\ref{crossingf} (a). The superhorizon ``cross out'' point corresponds to the 
point in Fig.~\ref{crossingf} (b). 
}
\label{schematicf}
\end{figure} 

To illustrate our discussions, we schematically show in Fig.~\ref{schematicf} the particle-antiparticle pair symmetry and asymmetry in different epochs and indicate the occurrences of subhorizon cross-in and superhorizon cross-out. 
At the end of ${\mathcal M}$-episode of reheating, a tiny asymmetry (\ref{net}) of massive particles and antiparticles appears inside the horizon. Such a nonzero net number of massive particles and antiparticles inside the 
horizon is preserved during the entire history of standard cosmology, since
superhorizon criteria $H^{-1}/\omega_\delta^{-1}= (2\Gamma_M/H)^{1/2} < 2$ (\ref{routside}) and $H>\Gamma_M$ holds after reheating. 

\section
{\bf Baryogenesis 
in reheating epoch}\label{bary}

We will show that the asymmetry of massive particles $X$ and antiparticles $\bar X$ 
of the net number density (\ref{net}) results in the asymmetry of baryon numbers of baryons $B$ and antibaryons $\bar B$ at the reheating end, which
leads to baryogenesis in agreement with observations. 

\subsection{Initial baryon asymmetry for standard cosmology}

In the reheating ${\mathcal R}$-episode 
after the superhorizon crossing (\ref{rhc}), unstable massive pairs rapidly decay into relativistic particles, and the final states are gauge bosons, leptons and baryons in the SM and dark matter particles. Massive pair decay process $\bar X X\Rightarrow \bar \ell\	\ell$ and all other 
microscopic processes respect the CPT symmetry and other fundamental symmetries. 
The asymmetry (\ref{net}) of 
massive particles $X$ and antiparticles $\bar X$ leads to the asymmetry between baryons and anti-baryons, the asymmetry between leptons and anti-leptons, and the asymmetry between dark-matter particles and antiparticles. These SM and dark-matter 
particle-antiparticle asymmetries are initial conditions for starting standard cosmology.
The asymmetry (\ref{net}) of $X$ and $\bar X$ also accounts for dark-matter particle and anti-particle asymmetry in the present Universe.

Such a scenario sharply contrasts with the usual dynamic scenarios that initial particles $X$ and antiparticles 
$\bar X$ compositions are symmetric. 
The production of baryon $B$ and anti-baryon $\bar B$ asymmetry requires the interactions between $\bar X, X$ and  $\bar B, B$ obey three Sakharov conditions that $B$- and $CP$-violating $X$ decays to baryons $B$ and decouples from thermal equilibrium in cosmic evolution, as discussed in the introduction. The particles $X$ and antiparticles 
$\bar X$ asymmetry (\ref{rhc}) created by the superhorizon crossing (\ref{net})  acts as an explicit $CP$-symmetry breaking between particles $X$ and antiparticles $\bar X$ in either their effective Lagrangian or their initial condition of compositions, which persists in standard cosmology after the reheating. This article will focus
only on the resultant asymmetry between the baryon and anti-baryon. The lepton-antilepton and dark-matter particle-antiparticle asymmetries will be issues for future studies.

\comment{
Such a nonzero net number of particles and antiparticles inside the 
horizon is preserved in the entire history of standard cosmology because there is
no another subhorizon crossing of the asymmetric mode $\delta_{M}$ after the superhorizon 
crossing (\ref{rhc}).  The reasons are the following. As the horizon scale $H$ decreases, 
the spacetime produced pairs have a small density $\rho^H_{_M}=2\chi m^2 H^2$ and the horizon 
$H$ evolution is governed by the matter produced in the reheating.
They rapidly decay to relativistic particles for $\tau_{_R}\ll \tau_{_M}$, 
and the oscillation ${\mathcal S} \Leftrightarrow \bar X X$ 
is no longer relevant. Therefore, the super horizon condition 
$H^{-1}/\omega_\delta <2$ (\ref{routside}) holds. The asymmetric modes $\delta_{M}$
carried the net particle number has been frozen 
in the superhorizon, and not reentered back to the horizon again.
, 
thus the net particle number appears, viewed by the subhorizon observer, in the standard cosmology. This net number density (\ref{net}) of massive particles and antiparticles 
is the initial value of the asymmetric (net) numbers of particles and antiparticles 
in standard cosmology. }

\subsection{Net baryon numbers and baryon number-to-entropy ratio}
 
Following the approach discussed in Ref.~\cite{Kolb1990}, we use the net particle number density $\delta n^{\rm crout}_{_M}$ (\ref{net}) and
the continuity equation of the net baryon number density 
$n_{_B}$ to obtain
\begin{eqnarray}
\dot n_{_B}+ 3 Hn_{_B} &=& \delta n^{\rm crout}_{_M}/\tau_{_R},\nonumber\\
\Rightarrow n_{_B}(a) &=&  2.31\times 10^{-4}n^H_{_M}(a_{_R})\left(\frac{a}{a_{_R}}\right)^{-3}[1-\exp - t/\tau_{_R}].
\label{brates}
\end{eqnarray}
Note that the net number density $\delta n^{\rm crout}_{_M}$ (\ref{net}) of massive particles and antiparticles is the counterpart of mean net baryon number density $\epsilon_{_{\rm CP}}n_{_B}$ in Ref.~\cite{Kolb1990}, attributing to an explicit $CP$-symmetry violating interaction in an effective Lagrangian. There, massive particles $X$ and antiparticles $\bar X$ $CP$-violating decays produce a mean net baryon number $\epsilon_{_{\rm CP}}n_{_B}$. Here, 
the superhorizon crossing of massive particles and antiparticles oscillating mode leads to the net number density $\delta n^{\rm crout}_{_M}$ (\ref{net}). Via massive particles' decay to baryons, the net number density $\delta n^{\rm crout}_{_M}$ leads to a net baryon number density in the right-handed side of the continuity equation (\ref{brates}).

In the second line of integration, the initial moment $t_i\ll \tau_{_R}$ is assumed and the solution 
for massive pairs
$\rho_{_M}$ decay is used, 
see Eq.~(7.23) of Ref.~\cite{Xue2023a} or 
the first equation in (6.61) of Ref.~\cite{Kolb1990}.
The physical content is clear: at late times, 
$t\gg \tau_{_R}$, the 
net baryon number per comoving volume $a^3n_{_B}(a)$ is just 
$2.31\times 10^{-4}$ times 
the initial number of massive particles per comoving volume 
$a^3_{_R}n^H_{_M}(a_{_R})$. Note that $a^3n_{_B}(a)$ is the comoving number 
density, whereas $n_{_B}(a)$ is the physical number density, i.e., 
the net baryon number per physical volume. Since the decay time scale 
$\tau_{_R}\propto \hat m^{-1}$ is very short, we adopt the approximation 
that the superhorizon crossing coincides 
with the genuine reheating (\ref{rhc}) 
($a=a_{_R}\approx a_{\rm crout}$ and $t\approx \tau_{_R}$) 
to obtain the net baryon number density
\begin{eqnarray}
n^R_{_B}(a_{_R}) &=&  1.46\times 10^{-4} n^H_{_M}(a_{_R}),
\quad n^H_{_M}(a_{_R})=\chi \hat m H_{\rm RH}^2.
\label{brates1}
\end{eqnarray}
It yields the origin of the net number of baryons or antibaryons, i.e., 
the baryogenesis in the Universe.

As a $CP$-asymmetric relic from the reheating epoch or as a $CP$-initial asymmetric 
state of the standard cosmology, the $CP$-asymmetry persists after the reheating. The net baryon number (\ref{brates1}) 
remains inside the horizon and conserves in subsequent 
Universe evolution. It 
accounts for the baryon and anti-baryon asymmetry observed today. 
It should be emphasized that in our scenario, the 
standard cosmology starts from the initial state of asymmetrical 
baryon and antibaryon numbers. Therefore the three Sakharov 
conditions \cite{Sakharov1967} of dynamical solution for the baryogenesis are not applicable.


Using the entropy $S_R$ and temperature $T_{\rm RH}$ of the reheating epoch, obtained in Eqs.~(8.6) and (8.8) of Ref.~\cite{Xue2023a},
we calculate the baryon asymmetry represented by the ratio of the net baryon 
number $a^3_{_R}n^R_{_B}$ (\ref{brates1}) to the entropy $S_R$. 
Per comoving volume, this ratio at the reheating $a_{_R}$ is given by  
\begin{eqnarray}
\frac{n^R_{_B}}{s_{_R}}=\frac{a^3_{_R}n^R_{_B}}{S_R}\approx   1.46\times 10^{-4}\frac{a^3_{_R}n^H_{_M}(a_{_R})}{S_R}
\approx 2.6 \times 10^{-4} \frac{T_{\rm RH}}{\hat m},
\label{rehtf}
\end{eqnarray}
where the entropy density $s_{_R}=S_R/a_{_R}^3$ at the genuine reheating ${\mathcal R}$-episode. 
This result is consistent with the one derived from the processes of 
baryon-number violating decay in Ref.~\cite{Kolb1990}. This 
baryon number-to-entropy ratio $n^R_{_B}/s_{_R}$ (\ref{rehtf}) preserves its value from the reheating epoch to the present time. The present observational value is  
$n_{_B}/s=  0.864^{+0.016}_{-0.015}\times 10^{-10}$ \cite{Ade2016}. 
This determines the ratio of the reheating temperature $T_{\rm RH}$ to 
the mass parameter $\hat m$,
\begin{eqnarray}
(T_{\rm RH}/\hat m)\approx 3.3\times 10^{-7}.
\label{tmr}
\end{eqnarray}
It is used to constrain the parameters $(g_Y^2/g^{1/2}_*)$ and 
$(\hat m/M_{\rm pl})$, 
see Eq.~(8.11) of Ref.~\cite{Xue2023a}, $g_*$ is the effective degeneracy of SM light particles $\bar \ell \ell$ produced in reheating.

In this article, we determine the ratio of parameters (\ref{tmr}) by considering the observed baryogenesis resulting from the asymmetry of particles $X$ and antiparticles $\bar X$ due to the horizon-crossing dynamics in the reheating epoch. 
Such an asymmetry can also cause SM leptogenesis 
and the asymmetry between sterile dark matter particles and antiparticles because particles $X$ and antiparticles $\bar X$ can annihilate and decay to SM leptons and dark matter particles. The ratio (\ref{tmr}) will be better constrained if leptogenesis and asymmetries of sterile dark matter particles and antiparticles are determined. These are subjects for future studies.


\section
{\bf Dark-matter acoustic wave and large-scale structure}\label{soundw}

We describe in Sec.~\ref{oscillationh} the density perturbation modes $\delta^{\bf k}_{_M}$ and $\Delta^{\bf k}_{_M}$ of massive particle-antiparticle pair plasma. We focus on  
the ``zero mode'' $|{\bf k}|=0$ oscillating modes 
$\delta^0_{_M}$ and $\Delta^0_{_M}$ of the frequency 
$\omega = (2H\Gamma)^{1/2}$ in Sec.~\ref{mode0}.  
We study their subhorizon crossing in Sec.~\ref{hcross0sub} and superhorizon crossing in Sec.~\ref{lcross}. The latter accounts 
for the baryogenesis 
in Sec.~\ref{bary}. 

From this section, we will study wave-propagating modes 
$\delta^{\bf k}_{_M}$ and $\Delta^{\bf k}_{_M}$ ($|{\bf k}|\not=0$) of the frequencies 
$\omega_{\delta,\Delta}(|{\bf k}|)$ (\ref{fdp+}) and 
(\ref{fdp-}). They represent dark-matter acoustic waves, as indicated by the red line in Fig.~\ref{schematicf}, that exited 
from (superhorizon crossing out) 
the horizon and reentered into (subhorizon crossing in) the horizon again. 
In particular, we focus on the pair symmetric density perturbation 
modes $\Delta^{\bf k}_{_M}$. We examine 
the negative term $4\pi G\rho^H_{_M}\Delta^{\bf k}_{_M}$ in the frequency 
$\omega_{\Delta}(|{\bf k}|)$ (\ref{fdp-}) 
to study the Jeans instability for the cases: 
(a) the pre-inflationary epoch where the mass parameter $m\gtrsim H$ 
and $v_s^2\lesssim 1$, 
namely, pairs can be relativistic; (b) the inflationary epoch where the mass parameter $m\gg H$ 
and $v_s^2\ll 1$, namely pairs are non-relativistic.
Here the mass parameter $m$ are the values for the 
pre-inflation or inflation epoch.
To be specific,
we investigate the wave-propagating modes $\Delta^{\bf k}_{_M}$ and $\delta^{\bf k}_{_M}$: 
\begin{enumerate}[(i)]
\item in the pre-inflation and inflation epochs, their superhorizon crossings out 
and modes frozen or amplified outside of the horizon; 
\item  after the recombination, their subhorizon crossings reentry the horizon, yielding 
peculiar ``dark-matter'' acoustic waves imprinting on the matter power spectrum 
at large length scales;
\item their possible impacts on forming large-scale structures and galaxy profiles.
\end{enumerate}

To start with, we would like to emphasize that the perturbations discussed here 
are the kinds of acoustic waves of the dispersion relations (\ref{fdp+}) 
and (\ref{fdp-}). They are
due to fluctuations in the form of the local equation of the state 
$\omega_{_M}= v^2_s$ (\ref{soundv}) of particles and antiparticles, namely, spatial 
fluctuations in the number of particles or antiparticles (compositions) 
per comoving volume. 
Its physical nature differs from the curvature perturbations as fluctuations 
in energy density characterized by the local value of the spatial curvature of the spacetime.

\subsection{Wave modes, horizon crossings, and Jeans instability}

The wave modes are described by the comoving
momentum $|{\bf k}|$ (wavelength $\lambda=|{\bf k}|^{-1}$), 
whose value is constant in time. As a reference, we consider the pivot 
scale $k^*=0.05 ({\rm Mpc})^{-1}$, at which the curvature perturbations cross 
outside and inside horizons twice, accounting for the CMB observations. The 
corresponding values of the horizon $k^*=(Ha)_*$, the $\epsilon$-rate
$\epsilon=\epsilon^*$ and the mass parameter
$m=m_*$ for inflation, that are given in Sec.~6 of Ref.~\cite{Xue2023} or \cite{Xue2023a}. 

The case $|{\bf k}|> k^*$ corresponds to the wave modes that exit the horizon in the 
pre-inflation epoch and reenter the horizon 
after the recombination epoch. 
The case $|{\bf k}| < k^*$ corresponds to the wave modes that exit the horizon in the 
inflation epoch and reenter the horizon before the recombination epoch. In both cases,
$H\gg \Gamma_M =(\chi/4\pi) m \epsilon$ and $\epsilon\ll 1$. 
Wave equations (\ref{kdp+}) and (\ref{kdp-}) approximately become,
\begin{eqnarray}
\ddot\delta^{\bf k}_{_M} +2H\dot\delta^{\bf k}_{_M} 
 &\approx & -\omega^2_{\delta}(|{\bf k}|) \delta^{\bf k}_{_M}, 
\label{dp+1}\\
\ddot\Delta^{\bf k}_{_M} +2H\dot\Delta^{\bf k}_{_M} 
&\approx & -\omega^2_{\Delta}(|{\bf k}|)\Delta^{\bf k}_{_M},
\label{dp-1}
\end{eqnarray}
where the Hubble rate $H$ slowly varies $H\approx {\rm const}$.
Following the same discussions in previous Sec.~\ref{mode0} or the usual discussions
of the curvature perturbations, these wave modes $\delta^{\bf k}_{_M}$ 
and $\Delta^{\bf k}_{_M}$ 
superhorizon crossing at $a^*$ when
$\omega_{\delta,\Delta}(|{\bf k}|)= 2H$ (\ref{hcri}), see Fig.~\ref{schematicf}. We use the frequencies $\omega_{\delta,\Delta}(|{\bf k}|)$ (\ref{fdp+}) and (\ref{fdp-}) to obtain the wavenumber $|{\bf k}|_{\delta,\Delta}$ at the superhorizon crossing,
\begin{eqnarray}
|{\bf k}|^2_{\delta \rm cross}&=&\Big(4-\chi\frac{\epsilon m}{2\pi H}\Big)\frac{(Ha)^2}{v_s^2}\Big|_{\rm cross}\approx 4\frac{(Ha)^2}{v_s^2}\Big|_{\rm cross},
\label{kmode-}\\
|{\bf k}|^2_{\Delta \rm cross}&=&\Big[4-\chi\frac{\epsilon m}{2\pi H}+\chi\Big(\frac{m}{m_{\rm pl}}\Big)^2\Big]\frac{(Ha)^2}{v_s^2}\Big|_{\rm cross}\nonumber\\
&\approx& 4\frac{(Ha)^2}{v_s^2}\Big|_{\rm cross}\approx |{\bf k}|^2_{\delta \rm cross}.
\label{kmode+}
\end{eqnarray}
The sound velocity $v_s$ value (\ref{soundv}) decreases from the relativistic case $v_s\lesssim 1$ (pre-inflation $m\gtrsim H$) to the non-relativistic case $v_s\ll 1$ (inflation $m\gg H$), as the horizon $H$ decreases and
the equation of state $p^H_{_M}=\omega^H_{_M}
\rho^H_{_M}$ (\ref{apdenm}) changes from $\omega^H_{_M}\lesssim 1/3$ to $\omega^H_{_M}\approx 0$ \cite{Xue2020,Xue2023}. 

The horizon crossing (\ref{kmode-}) and (\ref{kmode+}) 
shows that because of $v_s^2<1$, 
the modes $|{\bf k}|_{\delta {\rm cross},\Delta {\rm cross}} >k^*$ 
cross the horizon before the inflation pivot scale 
$k^*=(Ha)_*$, and reenter the horizon after the recombination $(Ha)_*$.
Therefore, these acoustic wave perturbations can cross to 
outside the horizon in the pre-inflation epoch and then cross inside the horizon 
after the recombination.  
The acoustic wave modes of larger 
$|{\bf k}|_{\delta {\rm cross},\Delta {\rm cross}}$ 
exit the horizon earlier and reenter the horizon later. They should leave their imprints on the linear regime of large-scale structure and the nonlinear regime of galaxy clustering. We called these acoustic or quantum wave modes of ``dark-matter'' waves \cite{Xue2023, Xue2020}, and will discuss them in detail below.


The particle-antiparticle symmetric modes $\Delta^{\bf k}_{_M}$ (\ref{delta2}) represent the acoustic waves of 
pair-density perturbations of the dispersion relation (\ref{fdp-}). 
It is analogous to the matter density perturbation 
wave in gravitation fields. 
We examine the dispersion relation (\ref{fdp-}) 
to check whether the Jeans instability occurs 
and the $\Delta^{\bf k}_{_M}$ amplitudes amplify for imaginary frequencies 
$\omega^2_{\Delta}(|{\bf k}|)<0$, i.e., 
\begin{eqnarray}
|{\bf k}|^2_{\Delta}&<&|{\bf k}|^2_{\rm Jeans}	\equiv \chi\Big[\Big(\frac{m}{m_{\rm pl}}\Big)^2-\frac{\epsilon m}{2\pi H}\Big]\frac{(Ha)^2}{v_s^2}.
\label{jeans}
\end{eqnarray} 
The $\omega^2_{\Delta}(|{\bf k}|)=0$ gives the Jeans wavenumber $|{\bf k}|_{\rm Jeans}$ and wavelength 
$\lambda_{\rm Jeans}=|{\bf k}|^{-1}_{\rm Jeans}$. The modes 
$\Delta^{\bf k}_{_M}$ of wavelengths $\lambda >\lambda_{\rm Jeans}$ undergo the Jeans instability 
because the gravitational attraction of pair-density perturbations prevails.

In the inflation epoch, the gravitational attractive term $4\pi G \rho^H_{_M}$ 
is negligible compared with the the 
``quasi mass'' term $2H\Gamma_M$ in the frequency 
$\omega^2_{\Delta}$ (\ref{fdp-}). Thus 
$\omega^2_{\Delta}>0$ and the Jeans instability does not occur. 
In the pre-inflation epoch, the term $2H\Gamma_M$ is negligible 
compared with the gravitational attractive term $4\pi G \rho^H_{_M}$.  
It is then possible to have the imaginary frequency (\ref{fdp+}) 
$\omega^2_{\Delta}< 0$ for 
$|{\bf k}|^2_{\Delta}<|{\bf k}|^2_{\rm Jeans}$ (\ref{jeans}), and
\begin{eqnarray}
|{\bf k}|^2_{\rm Jeans}	
&\approx& \chi\Big(\frac{m}{m_{\rm pl}}\Big)^2\frac{(Ha)^2}{v_s^2};\quad \chi\Big(\frac{m}{m_{\rm pl}}\Big)^2 <1,
\label{jeans1}
\end{eqnarray}
where the Jeans instability occurs. 

\subsubsection{Subhorizon stable modes and dark-matter acoustic waves}\label{stable}

For short-wavelength modes 
$|{\bf k}|^2_{\Delta}> |{\bf k}|^2_{\rm Jeans}$, 
the pressure terms $(v_s^2|{\bf k}|^2/a^2)$ are dominant in the frequency 
$\omega^2_{\Delta}$ (\ref{fdp-}), 
compared with the gravitational attraction $4\pi G \rho^H_{_M}$ 
of pair-density perturbations.  
Equation (\ref{dp-1}) for the pair-density perturbation approximately becomes  
\begin{eqnarray}
\ddot\Delta^{\bf k}_{_M} +2H\dot\Delta^{\bf k}_{_M} 
 &=& -(v_s^2|{\bf k}|^2/a^2)\Delta^{\bf k}_{_M},
\label{sound1}
\end{eqnarray}
which is the typical Mukhanov-Sasaki equation, similar to the one for 
the curvature perturbation. 
Using Eqs.~(\ref{hosi}), (\ref{hosid}) and (\ref{hosis}), we find its solution
\begin{eqnarray}
\Delta^{\bf k}_{_M}(t)\propto e^{ -Ht}\exp - i\tilde\omega_{\Delta}(|{\bf k}|)(1-\zeta^2)^{1/2}t, 
\quad \zeta \approx \frac{H}{\tilde\omega_{\Delta}(|{\bf k}|)}= \frac{Ha}{v_s|{\bf k}|_{\Delta}},
\label{sounds}
\end{eqnarray}
where the acoustic wave frequency $\tilde\omega_{\Delta}(|{\bf k}|)= v_s|{\bf k}|_{\Delta}/a$. 

For given comoving horizon $(Ha)$ and sound velocity $v_s$,
horizon crossing wavenumber $|{\bf k}|_{\Delta \rm cross}=2Ha/v_s$ (\ref{kmode+})
is larger than Jeans wavenumber $|{\bf k}|_{\rm Jeans}$ (\ref{jeans1}),
\begin{eqnarray}
|{\bf k}|_{\Delta \rm cross}>|{\bf k}|_{\rm Jeans}.
\label{candj}
\end{eqnarray}
This shows that the pair-density perturbations $\Delta^{\bf k}_{_M}$ 
oscillate as 
\begin{enumerate}[(i)]

\item  {\it  sub-horizon sized modes}: an underdamped acoustic wave inside the 
horizon for $\zeta<1$ and $|{\bf k}|_{\Delta} > |{\bf k}|_{\Delta \rm cross}>|{\bf k}|_{\rm Jeans}$.
The modes $\Delta^{\bf k}_{_M}(t)$ are stable acoustic waves;

\item  {\it  super-horizon sized modes}: an overdamped acoustic wave outside the horizon for $\zeta>1$ 
and $|{\bf k}|_{\Delta \rm cross} > |{\bf k}|_{\Delta} >|{\bf k}|_{\rm Jeans}$. The mode amplitudes
$\Delta^{\bf k}_{_M}\propto {\rm const}$ are frozen.
\end{enumerate}
On the other hand, as the comoving horizon $(Ha)$ increases and the sound 
velocity $v_s$ decreases in the pre-inflation and inflation epochs, 
the horizon crossing wavenumber $|{\bf k}|_{\Delta \rm cross}$ (\ref{kmode+})
increases.
The acoustic wave modes $\Delta^{\bf k}_{_M}$ of fixed wavenumber $|{\bf k}|_{\Delta}$ 
evolve from {\it  subhorizon}
to {\it superhorizon}, and 
the horizon crossing occurs at $|{\bf k}|_{\Delta}=|{\bf k}|_{\Delta \rm cross}$. 

Moreover, the {\it  super-horizon sized mode} $|{\bf k}|_{\Delta}$ 
reenters the horizon at $(Ha)_{\rm reenter}$ and becomes 
a {\it  sub-horizon sized mode}, when 
\begin{eqnarray}
|{\bf k}|_{\Delta}=(Ha)_{\rm reenter} ,
\label{reenter}
\end{eqnarray}
where $(Ha)^{-1}_{\rm reenter}$ is the comoving horizon size after 
the recombination. Such mode $|{\bf k}|_{\Delta}=({Ha})_{\rm reenter}$ 
behaves as an acoustic wave of dark-matter 
density perturbations. It imprints on the matter power spectrum of low-$\ell$ 
multipoles $\ell\leq 2$, corresponding to large length scales. 
It is reminiscent of baryon acoustic oscillations due to the 
coupling in the baryon-photon fluid.

These are qualitative discussions on dark-matter acoustic waves, possibly relevant for observations. However, the 
quantitative results depend not only 
on the initial amplitude value $\Delta^{\bf k}_{_M}(0)$, 
the wavenumber $|{\bf k}|_{\Delta}$ 
and the horizon crossing size $(Ha)_{\rm cross}$, but also on the sound velocity $v_s$ 
in Eq.~(\ref{sounds}).

\subsubsection{Unstable superhorizon modes and large-scale structure}

For long-wavelength modes $|{\bf k}|^2_{\Delta}\ll |{\bf k}|^2_{\rm Jeans}$, 
the pressure terms $(v_s^2|{\bf k}|^2/a^2)$ are negligible in the frequency 
$\omega^2_{\Delta}$ (\ref{fdp-}), 
compared with the gravitational attraction $4\pi G \rho^H_{_M}$ of pair-density perturbations. 
Equation (\ref{dp-1}) for the pair-density perturbation approximately becomes    
\begin{eqnarray}
\ddot\Delta^{\bf k}_{_M} +2H\dot\Delta^{\bf k}_{_M} 
 &=& \chi\Big(\frac{m}{m_{\rm pl}}\Big)^2 H^2\Delta^{\bf k}_{_M},
\label{kdp--}
\end{eqnarray}
as the microscopic physics (e.g., pressure terms $(v_s^2|{\bf k}|^2/a^2)$) is impotent and negligible. The Hubble rate $H$ is 
approximately a constant, slowly varies in the pre-inflation and inflation epochs. 

Equation (\ref{kdp--}) is a new kind of differential equation for density perturbations, 
differently from Eq.~(\ref{sound1}) of the Mukhanov-Sasaki type.
This equation (\ref{kdp--}) has two independent solutions: 
\begin{eqnarray}
\Delta^{\bf k}_{_M}(t)\propto \exp-2Ht; \quad \Delta^{\bf k}_{_M}(t)\propto \exp +\frac{\chi H}{2}\Big(\frac{m}{m_{\rm pl}}\Big)^2t .
\label{ampli}
\end{eqnarray}
At late times, the exponentially glowing modes of pair-density perturbations are crucial. Whereas
the exponentially decaying modes physically correspond to those modes with initial overdensity and
velocity arranged so that the initial velocity perturbation eventually 
eases pair-density perturbations. 

Equation (\ref{jeans1}) shows that $|{\bf k}|^2_{\Delta}\ll |{\bf k}|^2_{\rm Jeans}$  
means $|{\bf k}|^2_{\Delta}< |{\bf k}|^2_{\Delta \rm cross}$, 
indicating these modes (\ref{ampli}) are superhorizon size.
It is consistent with neglecting the pressure term $(v_s^2|{\bf k}|^2/a^2)$ 
of the microscopic physics that cannot causally arrange 
the pair-density perturbations in superhorizon size. 
As a result, the pressure terms do not balance the gravitational attraction, 
leading to an increase in the amplitudes of pair-density perturbations. 
These unstable and superhorizon-sized modes (\ref{ampli}) exponentially 
glow in time. Its characteristic time scale is given by the second solution 
in Eq.~(\ref{ampli})
\begin{eqnarray}
\tau^{-1}_{_\Delta} = \frac{\chi H}{2}\Big(\frac{m}{m_{\rm pl}}\Big)^2.
\label{amplit}
\end{eqnarray}
These solutions (modes)  
differ from the superhorizon-sized modes of frozen constant 
amplitudes (\ref{hosis}) or (\ref{sounds}) for $\zeta \gg 1$. 

We further 
study such an unstable and superhorizon-sized mode of fixed wavenumber 
$|{\bf k}|_{\Delta}$ in the range
\begin{eqnarray}
|{\bf k}|_{\rm Jeans}>|{\bf k}|_{\Delta} > k^*,
\label{amplic}
\end{eqnarray}
where $|{\bf k}|_{\Delta {\rm cross}} > |{\bf k}|_{\rm Jeans}$ (\ref{candj}) 
and  $k^*$ is the inflation pivot scale of CMB  observations. The initial amplitudes 
$\Delta^{\bf k}_{_M}(0)$ of such modes (\ref{ampli}) and (\ref{amplic}) are very small, 
as the curvature perturbations. However, they could greatly amplify in the superhorizon before reentering the subhorizon. Therefore, 
it is possible that such modes $\Delta^{\bf k}_{_M}$ of 
the pair-density perturbation
are no longer a small perturbation, when they reenter the horizon and their
wavelength $|{\bf k}|^{-1}_{\Delta}$
becomes subhorizon-sized, 
$|{\bf k}|_{\Delta} =(Ha)_{\rm reenter}$ (\ref{kdp--}). The subhorizon crossing occurs after the recombination
$(Ha)_{\rm reenter} < (Ha)_*$.  
As a consequence, this phenomenon could play crucial physical roles 
in the formations of large-scale structures and galaxy cluster profiles at much larger scales.

We recall the basic scenario of primordial curvature perturbations
leading to the large-scale structure in the standard cosmology. 
The curvature perturbations, whose amplitudes are small constants in superhorizon, 
reenter the horizon and 
lead to the CMB temperature anisotropic fluctuation
$\delta T/T\sim {\mathcal O}(10^{-5})$. Such fluctuations
relate to the matter density perturbations 
$\delta \rho/\rho \propto \delta T/T$
at the recombination of the redshift $z\sim 10^{3}$. These matter density perturbations (amplitudes) are small, and their physical 
sizes (wavelengths) increase linearly as the scale factor $a(t)$. 
However, under the influence of their 
gravitational attractions and the Jeans 
instability, the matter density perturbations 
glow $\delta \rho/\rho \propto {\mathcal O}(1)$ and become
nonlinear, therefore approximately maintain  constant physical sizes, 
eventually forming a large-scale structure. 

Our qualitative analysis and 
discussions show the following additional possibilities.
The amplitudes of unstable pair-density perturbation modes in the range  (\ref{amplic}) get amplification in the superhorizon up to 
the order of unity $\Delta^{\bf k}_{_M}\propto {\mathcal O}(1)$, when they reenter the horizon after the
recombination. Therefore, they   
should have some physical consequences on the formation of large-scale structures 
and galaxies' profiles, as well as homogeneities on scales beyond $\sim 100Mpc/h$.
However, we are not able to give quantitative results in this article and 
further studies are required.  

\subsection{Particle-antiparticle ``neutral plasma'' acoustic wave}

We turn to the discussions of the particle-antiparticle density perturbations 
$\delta^{\bf k}_{_M}$, described by the frequency 
$\omega_\delta({\bf k})$ (\ref{fdp+}) and wave mode equation (\ref{dp+1}), 
\begin{eqnarray}
\ddot\delta^{\bf k}_{_M} +2H\dot\delta^{\bf k}_{_M} 
 &=& -(v_s^2|{\bf k}|^2/a^2)\delta^{\bf k}_{_M},
\label{sound1s}
\end{eqnarray}
and horizon crossing (\ref{kmode+}). 
These are the same equations as those for the pair-density 
perturbations $\Delta^{\bf k}_{_M}$, except the absence of 
the gravitational attraction term $4\pi G\rho^H_{_M}$ 
and Jeans instability. The solution is similar to the stable $\Delta^{\bf k}_{_M}$ 
modes (\ref{sounds}) 
\begin{eqnarray}
\delta^{\bf k}_{_M}(t)\propto e^{ -Ht}\exp - i\tilde\omega_{\delta}(|{\bf k}|)(1-\zeta^2)^{1/2}t, 
\quad \zeta \approx \frac{H}{\tilde\omega_{\delta}(|{\bf k}|)}= \frac{Ha}{v_s|{\bf k}|_{\delta}},
\label{soundss}
\end{eqnarray}
where the acoustic wave frequency 
$\tilde\omega_{\delta}(|{\bf k}|)= v_s|{\bf k}|_{\delta}/a$. 
For given comoving horizon $(Ha)$, sound velocity $v_s$,
and horizon crossing wavenumber $|{\bf k}|_{\delta {\rm cross}}=2Ha/v_s$ (\ref{kmode-}), 
the particle-antiparticle density perturbation modes $\delta^{\bf k}_{_M}$ behave as 
\begin{enumerate}[(i)]

\item  {\it  sub-horizon sized modes}: an underdamped acoustic wave mode inside the 
horizon for $\zeta<1$ and $|{\bf k}|_{\delta} > |{\bf k}|_{\delta {\rm cross}}$.
The modes $\delta^{\bf k}_{_M}(t)$ are stable acoustic wave modes;

\item  {\it  super-horizon sized modes}: an overdamped acoustic wave mode outside the horizon for $\zeta>1$ 
and $|{\bf k}|_{\delta} <|{\bf k}|_{\delta \rm cross}$. The mode amplitudes
$\delta^{\bf k}_{_M}\propto {\rm const}$ are frozen.

\end{enumerate}

On the other hand, as comoving horizon $(Ha)$ increases and sound 
velocity $v_s$ decreases in the pre-inflation epoch, 
horizon crossing wavenumber $|{\bf k}|_{\delta \rm cross}$ (\ref{kmode-})
increases.
The modes $\delta^{\bf k}_{_M}$ of fixed wavenumber $|{\bf k}|_{\delta}$ 
evolve from a {\it  sub-horizon sized mode}
to a {\it super-horizon sized mode}, and 
horizon crossing occurs at $|{\bf k}|_{\delta}=|{\bf k}|_{\delta \rm cross}$ (\ref{kmode-}). 
Moreover, {\it super-horizon sized mode} $|{\bf k}|_{\delta}$ 
reenters the horizon at the horizon $(Ha)_{\rm reenter}$, when
\begin{eqnarray}
|{\bf k}|_{\delta}=(Ha)_{\rm reenter},
\label{reenters}
\end{eqnarray}
after the recombination. It becomes a {\it sub-horizon sized mode} again. 

The modes $|{\bf k}|_{\delta}=({Ha})_{\rm reenter}$  of 
the particle-antiparticle density perturbations $\delta^{\bf k}_{_M}$ (\ref{delta2}) 
represent the acoustic waves of 
particle and antiparticle oscillations of the dispersion relation (\ref{fdp+}). They are
analogous to the neutral plasma oscillations of electrons and positrons. 
They respect the symmetry of particles and antiparticles.
These modes exit the horizon in the pre-inflation epoch and reenter the horizon 
after the recombination. It implies that such acoustic
waves from the primordial Universe would leave their imprints on the Universe 
after the recombination. It possibly imprints on the 
matter power spectrum 
of low-$\ell$ multipoles $\ell\leq 2$, corresponding to large-length scales.
However, we expect that the mode amplitudes 
$\delta^{\bf k}_{_M}$ should be small, given the 
energy densities $\rho^+_{_M}$ and $\rho^-_{_M}$ 
of particles and antiparticles are small. Namely, the pair energy densities 
$\rho^H_{_M}\approx \rho^+_{_M}+\rho^-_{_M}$ 
and $\rho^+_{_M}\approx \rho^-_{_M}$ 
are small in the pre-inflation epoch.  

In this section, we qualitatively describe three types of
``dark-matter'' acoustic waves originated from the particle-antiparticle 
oscillations in the pre-inflation epoch: (i) 
stable subhorizon pair-density perturbations; (ii) unstable superhorizon pair-density perturbations;
(iii) particle-antiparticle density perturbations. We 
present some discussions on their returns to the horizon after the recombination 
and possible relevance for observations. 
However, we cannot give quantitative results that depend not only 
on perturbation modes' wavenumbers $|{\bf k}|_{\Delta,\delta}$, 
initial amplitude values $\Delta^{\bf k}_{_M}(0)$ and $\delta^{\bf k}_{_M}(0)$,
and sound velocity $v_s$ 
in their oscillating equations (\ref{sound1}) and (\ref{sound1s}), but also on 
horizon crossing size $(Ha)^{-1}$  (\ref{kmode-}) and (\ref{kmode+}). We mention that the discussions here can be applied to dark-matter waves due to the quantum fluctuation of dark-matter particles.   
\comment{Remarks: gauge invariance when the modes of the curvature perturbation 
go to superhorizon. Namely, for superhorizon-sized modes, the gauge non-invariance 
of curvature perturbations. Instead, the density perturbations
$\delta^{\bf k}_{_M}$ and $\Delta^{\bf k}_{_M}$ are not gauge field mode, 
they are matter modes, what must be important and should be taken into account or
reckoned with; some fluctuation and effect 
of general relativity for $v_s\lesssim 1$, the Newtonian analysis has its
limitations} 

\section{Summary and remarks}\label{conclusion}

The article presents studies of particle-antiparticle asymmetry, baryogenesis 
and dark-matter acoustic waves in the scenario $\tilde\Lambda$CDM. We summarize the lengthy article by briefly summarizing
the fundamental equations adopted, basic physical phenomena described and results obtained that possibly give some insight into the issues of baryogenesis, dark matter particle-antiparticle asymmetry and large-scale structures of the Universe.
 
We study the perturbations of particle and antiparticle densities of massive pair plasma state in the reheating.
Starting from Eqs.~(\ref{continuum}-\ref{rho0}) for 
the density perturbations of particles and antiparticles, we derive the 
acoustic wave equations (\ref{kdp+}-\ref{kdp-}) for the particle-antiparticle
symmetric pair-density 
perturbation $\Delta_{_M}$ (\ref{delta1}) and 
the particle-antiparticle asymmetric 
density perturbation $\delta_{_M}$ (\ref{delta2}). 
We study the oscillating behaviours of the lowest-lying modes of zero 
wavenumbers $|{\bf k}|_{\Delta,\delta}=0$. We show that they evolve 
from superhorizon to subhorizon (\ref{ifinside}) in the preheating episode and
from subhorizon to superhorizon (\ref{routside}) in the genuine reheating episode. 
In the latter case, $\delta_{_M}$ undergoes horizon crossing to generate the net number density (\ref{net}) of 
particle and antiparticle, leading to the dark-matter particle-antiparticle asymmetry and
baryogenesis phenomenon. We obtain the baryon number-to-entropy ratio (\ref{rehtf}) consistent with observations. The $\delta_{_M}$ measures the relative (contrast) density of particles $X$ and antiparticles $\bar X$, 
whose spacetime oscillations are not in phase. This aspect is analogous to the isocurvature perturbation of the relative 
density of different particle species distributions, which are not in phase in spacetime. Here, we study the asymmetric perturbations $\delta_{_M}$, their super-horizon-crossing in the reheating epoch and never return to the horizon, which 
could account for the baryogenesis 
of the baryon number-to-entropy ratio $n^R_{_B}/s_{_R}$ observed in the present Universe. As a result, the ratio (\ref{tmr}) of reheating temperature $T_{\rm RH}$ and energy-mass scale $\hat m$ is constrained. Whereas, the isocurvature perturbation from the inflation epoch 
is due to the perturbation of scalar field 
components differing from 
the dominant scalar-field component that drives inflation. Their return to the horizon and impact on the last-scattering epoch can be constrained by the CMB observations.

In addition, we study the acoustic wave-propagating modes ($|{\bf k}|_{\Delta,\delta}\not=0$) 
of the pair-density perturbation $\Delta^{\bf k}_{_M}$ (\ref{dp-1}) and 
the particle-antiparticle asymmetric density perturbation $\delta^{	\bf k}_{_M}$ (\ref{dp+1}). They represent the dark-matter acoustic waves. We
show how they become superhorizon-sized modes (\ref{kmode-}) and (\ref{kmode+}) 
in the pre-inflation epoch, then return to the horizon 
$|{\bf k}|_{\Delta,\delta}=(Ha)_{\rm reenter}$ 
after the recombination. These modes can behave as stable acoustic waves 
of dark matter 
density perturbations. They possibly 
imprint on the matter power spectrum of low-$\ell$ multipoles $\ell\leq 2$, 
corresponding to large-length scales. 
Due to the Jeans instability of 
the pair-density perturbation $\Delta^{\bf k}_{_M}$ (\ref{kdp--}), the tiny amplitudes $\Delta^{\bf k}_{_M}\ll {\mathcal O}(1)$ of unstable superhorizon sized modes can get 
amplified. When they reenter the 
horizon after the recombination, these modes of 
pair-density perturbations can be of the order of unity $\Delta^{\bf k}_{_M}\propto {\mathcal O}(1)$. 
As a consequence, 
they should have some physical influences on forming large-scale structures and galaxy cluster profiles. 
It would be worthwhile to study 
whether these Jeans-amplified pair-density perturbations 
$\Delta^{\bf k}_{_M}$ produce primordial gravitational waves when they reenter the horizon. 

Further studies are required, and more elaborate numerical computations are needed.
Nevertheless, we expect that our theoretical scenario and results will provide valuable insights into baryogenesis, dark matter particle-antiparticle asymmetry and 
the role of dark matter acoustic waves, as well as their impacts on the evolution and structure of the Universe.
About the last point, it is worthwhile to mention the observational evidence indicating that dark matter inhomogeneities on scales beyond $\sim 100 {\rm Mpc}/h$ are larger than what is predicated by the theoretical $\Lambda$CDM model.

\comment{About the last point, it is worthwhile to mention that observational evidence of anomalously large structures is independently provided by the measurement of the impact of large dark matter fluctuations on the CMB via the integrated Sachs-Wolfe effect (ISW) (Flender et al. 2013, Flender S., Hotchkiss S., Nadathur S., 2013, JCAP, 2013, 013). The magnitude of the observed signal is more than $3\sigma$ larger than the theoretical $\Lambda$CDM expectation, indicating that dark matter inhomogeneities on scales beyond $\sim 100 {\rm Mpc}/h$ are larger than expected. Moreover, cosmic structures can
form bigger than this limit, displayed by a statistically significant clustering of the Gamma Ray Bursts sample at $1.6 < z < 2.1$ references to Fulvio Melia of https://arxiv.org/abs/2311.06249v1.  and https://www.aanda.org/articles/aa/abs/2014/01/aa23020-13/aa23020-13.html and https://academic.oup.com/mnras/article/452/3/2236/1078524 }



\begin{thebibliography}{10}

\bibitem{Starobinsky1980}
A.~A. Starobinsky, \emph{A new type of isotropic cosmological models without
  singularity}, \href{https://doi.org/10.1016/0370-2693(80)90670-X}{\emph{Phys.
  Lett. B} {\bfseries 91} (1980) 99}.

\bibitem{Guth1981}
A.~H. Guth, \emph{The inflationary universe: A possible solution to the horizon
  and flatness problems},
  \href{https://doi.org/10.1103/PhysRevD.23.347}{\emph{Phys. Rev. D} {\bfseries
  23} (1981) 347}.

\bibitem{Linde:1981mu}
A.~D. Linde, \emph{A new inflationary universe scenario: A possible solution of
  the horizon, flatness, homogeneity, isotropy and primordial monopole
  problems}, \href{https://doi.org/10.1016/0370-2693(82)91219-9}{\emph{Phys.
  Lett. B} {\bfseries 108} (1982) 389–393}.

\bibitem{Mukhanov:1982nu}
V.~F. Mukhanov and G.~V. Chibisov, \emph{The vacuum energy and large scale
  structure of the universe}, {\emph{Sov. Phys. JETP} {\bfseries 56} (1982)
  258–265}.

\bibitem{Albrecht1982}
A.~Albrecht and P.~J. Steinhardt, \emph{Cosmology for grand unified theories
  with radiatively induced symmetry breaking},
  \href{https://doi.org/10.1103/PhysRevLett.48.1220}{\emph{Phys. Rev. Lett.}
  {\bfseries 48} (1982) 1220}.

\bibitem{Linde1983}
A.~D. Linde, \emph{Chaotic inflation},
  \href{https://doi.org/10.1016/0370-2693(83)90837-7}{\emph{Phys. Lett. B}
  {\bfseries 129} (1983) 177}.

\bibitem{Kallosh:2021mnu}
R.~Kallosh and A.~Linde, \emph{BICEP/Keck and cosmological attractors},
  \href{https://doi.org/10.1088/1475- 7516/2021/12/008}{\emph{JCAP} {\bfseries
  12} (2021) 008} [\href{https://arxiv.org/abs/2110.10902}{{\ttfamily
  2110.10902}}].

\bibitem{Kofman1994}
L.~Kofman, A.~D. Linde and A.~A. Starobinsky, \emph{Reheating after inflation},
  \href{https://doi.org/10.1103/PhysRevLett.73.3195}{\emph{Phys. Rev. Lett.}
  {\bfseries 73} (1994) 3195}
  [\href{https://arxiv.org/abs/hep-th/9405187}{{\ttfamily hep-th/9405187}}].

\bibitem{Kofman1997}
L.~Kofman, A.~D. Linde and A.~A. Starobinsky, \emph{Towards the theory of
  reheating after inflation},
  \href{https://doi.org/10.1103/PhysRevD.56.3258}{\emph{Phys. Rev. D}
  {\bfseries 56} (1997) 3258}
  [\href{https://arxiv.org/abs/hep-ph/9704452}{{\ttfamily hep-ph/9704452}}].

\bibitem{Shtanov1995}
Y.~Shtanov, J.~H. Traschen and R.~H. Brandenberger, \emph{Universe reheating
  after inflation}, \href{https://doi.org/10.1103/PhysRevD.51.5438}{\emph{Phys.
  Rev. D} {\bfseries 51} (1995) 5438}
  [\href{https://arxiv.org/abs/hep-ph/9407247}{{\ttfamily hep-ph/9407247}}].

\bibitem{Bassett1998}
B.~A. Bassett and S.~Liberati, \emph{Geometric reheating after inflation},
  \href{https://doi.org/10.1103/PhysRevD.60.049902}{\emph{Phys. Rev. D}
  {\bfseries 58} (1998) 021302}
  [\href{https://arxiv.org/abs/hep-ph/9709417}{{\ttfamily hep-ph/9709417}}].

\bibitem{Tsujikawa1999}
S.~Tsujikawa, K.-i. Maeda and T.~Torii, \emph{Resonant particle production with
  nonminimally coupled scalar fields in preheating after inflation},
  \href{https://doi.org/10.1103/PhysRevD.60.063515}{\emph{Phys. Rev. D}
  {\bfseries 60} (1999) 063515}
  [\href{https://arxiv.org/abs/hep-ph/9901306}{{\ttfamily hep-ph/9901306}}].

\bibitem{Podolsky2002}
D.~I. Podolsky and A.~A. Starobinsky, \emph{Chaotic reheating}, {\emph{Grav.
  Cosmol. Suppl.} {\bfseries 8N1} (2002) 13}
  [\href{https://arxiv.org/abs/astro-ph/0204327}{{\ttfamily
  astro-ph/0204327}}].

\bibitem{Allahverdi2010}
R.~Allahverdi, R.~Brandenberger, F.-Y. Cyr-Racine and A.~Mazumdar,
  \emph{Reheating in inflationary cosmology: Theory and applications},
  \href{https://doi.org/10.1146/annurev.nucl.012809.104511}{\emph{Ann. Rev.
  Nucl. Part. Sci.} {\bfseries 60} (2010) 27}
  [\href{https://arxiv.org/abs/1001.2600}{{\ttfamily 1001.2600}}].

\bibitem{Amin2012}
M.~A. Amin, R.~Easther, H.~Finkel, R.~Flauger and M.~P. Hertzberg,
  \emph{Oscillons after inflation},
  \href{https://doi.org/10.1103/PhysRevLett.108.241302}{\emph{Phys. Rev. Lett.}
  {\bfseries 108} (2012) 241302}
  [\href{https://arxiv.org/abs/1106.3335}{{\ttfamily 1106.3335}}].

\bibitem{Amin2014}
M.~A. Amin, M.~P. Hertzberg, D.~I. Kaiser and J.~Karouby, \emph{Nonperturbative
  dynamics of reheating after inflation: A review},
  \href{https://doi.org/10.1142/S0218271815300037}{\emph{Int. J. Mod. Phys. D}
  {\bfseries 24} (2014) 1530003}
  [\href{https://arxiv.org/abs/1410.3808}{{\ttfamily 1410.3808}}].

\bibitem{Adshead2020}
P.~Adshead, J.~T. Giblin, M.~Pieroni and Z.~J. Weiner, \emph{Constraining axion
  inflation with gravitational waves across 29 decades in frequency},
  \href{https://doi.org/10.1103/PhysRevLett.124.171301}{\emph{Phys. Rev. Lett.}
  {\bfseries 124} (2020) 171301}
  [\href{https://arxiv.org/abs/1909.12843}{{\ttfamily 1909.12843}}].

\bibitem{Parker1973}
L.~Parker and S.~A. Fulling, \emph{Quantized matter fields and the avoidance of
  singularities in general relativity},
  \href{https://doi.org/10.1103/PhysRevD.7.2357}{\emph{Phys. Rev. D} {\bfseries
  7} (1973) 2357}.

\bibitem{Starobinsky1982}
A.~A. Starobinsky, \emph{Nonsingular model of the universe with the
  quantum-gravitational de sitter stage and its observational consequences},
  {\emph{Proc. of the Second Seminar ``Quantum Theory of
  Gravity", Moscow, October 1981, INR Press, Moscow} (1982) 58}.

\bibitem{Ford1987}
L.~H. Ford, \emph{Gravitational particle creation and inflation},
  \href{https://doi.org/10.1103/PhysRevD.35.2955}{\emph{Phys. Rev. D}
  {\bfseries 35} (1987) 2955}.

\bibitem{Kolb1996}
E.~W. Kolb, A.~D. Linde and A.~Riotto, \emph{Gut baryogenesis after
  preheating}, \href{https://doi.org/10.1103/PhysRevLett.77.4290}{\emph{Phys.
  Rev. Lett.} {\bfseries 77} (1996) 4290}
  [\href{https://arxiv.org/abs/hep-ph/9606260}{{\ttfamily hep-ph/9606260}}].

\bibitem{Chung2001}
D.~J.~H. Chung, P.~Crotty, E.~W. Kolb and A.~Riotto, \emph{On the gravitational
  production of superheavy dark matter},
  \href{https://doi.org/10.1103/PhysRevD.64.043503}{\emph{Phys. Rev. D}
  {\bfseries 64} (2001) 043503}
  [\href{https://arxiv.org/abs/hep-ph/0104100}{{\ttfamily hep-ph/0104100}}].

\bibitem{Chung2000}
D.~J.~H. Chung, E.~W. Kolb, A.~Riotto and I.~I. Tkachev, \emph{Probing
  Planckian physics: Resonant production of particles during inflation and
  features in the primordial power spectrum},
  \href{https://doi.org/10.1103/PhysRevD.62.043508}{\emph{Phys. Rev. D}
  {\bfseries 62} (2000) 043508}
  [\href{https://arxiv.org/abs/hep-ph/9910437}{{\ttfamily hep-ph/9910437}}].

\bibitem{Chung2003}
D.~J.~H. Chung, \emph{Classical inflation field induced creation of superheavy
  dark matter}, \href{https://doi.org/10.1103/PhysRevD.67.083514}{\emph{Phys.
  Rev. D} {\bfseries 67} (2003) 083514}
  [\href{https://arxiv.org/abs/hep-ph/9809489}{{\ttfamily hep-ph/9809489}}].

\bibitem{Chung2005}
D.~J.~H. Chung, E.~W. Kolb, A.~Riotto and L.~Senatore, \emph{Isocurvature
  constraints on gravitationally produced superheavy dark matter},
  \href{https://doi.org/10.1103/PhysRevD.72.023511}{\emph{Phys. Rev. D}
  {\bfseries 72} (2005) 023511}
  [\href{https://arxiv.org/abs/astro-ph/0411468}{{\ttfamily
  astro-ph/0411468}}].

\bibitem{Chung2019}
D.~J.~H. Chung, E.~W. Kolb and A.~J. Long, \emph{Gravitational production of
  super-Hubble-mass particles: an analytic approach},
  \href{https://doi.org/10.1007/JHEP01(2019)189}{\emph{JHEP} {\bfseries 01}
  (2019) 189} [\href{https://arxiv.org/abs/1812.00211}{{\ttfamily
  1812.00211}}].

\bibitem{Ade2016}
{\scshape Planck} collaboration, \emph{Planck 2015 results. xiii. cosmological
  parameters}, \href{https://doi.org/10.1051/0004-6361/201525830}{\emph{Astron.
  Astrophys.} {\bfseries 594} (2016) A13}
  [\href{https://arxiv.org/abs/1502.01589}{{\ttfamily 1502.01589}}].

\bibitem{Sakharov1967}
A.~D. Sakharov, \emph{Violation of CP invariance, C asymmetry, and baryon
  asymmetry of the universe}, {\emph{Soviet Journal of Experimental and
  Theoretical Physics Letters} {\bfseries 5} (1967) 24}.

\bibitem{Dolgov1982}
A.~D. Dolgov and A.~D. Linde, \emph{Baryon asymmetry in inflationary universe},
  \href{https://doi.org/10.1016/0370-2693(82)90292-1}{\emph{Phys. Lett. B}
  {\bfseries 116} (1982) 329}.

\bibitem{Abbott1982}
L.~F. Abbott, E.~Farhi and M.~B. Wise, \emph{Particle production in the new
  inflationary cosmology},
  \href{https://doi.org/10.1016/0370-2693(82)90867-X}{\emph{Phys. Lett. B}
  {\bfseries 117} (1982) 29}.

\bibitem{Kuzmin1985}
V.~A. Kuzmin, V.~A. Rubakov and M.~E. Shaposhnikov, \emph{On the anomalous
  electroweak baryon number nonconservation in the early universe},
  \href{https://doi.org/10.1016/0370-2693(85)91028-7}{\emph{Phys. Lett. B}
  {\bfseries 155} (1985) 36}.

\bibitem{Dine2003}
M.~Dine and A.~Kusenko, \emph{The origin of the matter-antimatter asymmetry},
  \href{https://doi.org/10.1103/RevModPhys.76.1}{\emph{Rev. Mod. Phys.}
  {\bfseries 76} (2003) 1}
  [\href{https://arxiv.org/abs/hep-ph/0303065}{{\ttfamily hep-ph/0303065}}].

\bibitem{Trodden_1999}
M.~Trodden, \emph{Electroweak baryogenesis},
  \href{https://doi.org/10.1103/revmodphys.71.1463}{\emph{Reviews of Modern
  Physics} {\bfseries 71} (1999) 1463–1500}.

\bibitem{Morrissey_2012}
D.~E. Morrissey and M.~J. Ramsey-Musolf, \emph{Electroweak baryogenesis},
  \href{https://doi.org/10.1088/1367-2630/14/12/125003}{\emph{New Journal of
  Physics} {\bfseries 14} (2012) 125003}.

\bibitem{Dolgov1995}
A.~Dolgov and K.~Freese, \emph{Calculation of particle production by Nambu
  goldstone bosons with application to inflation reheating and baryogenesis},
  \href{https://doi.org/10.1103/PhysRevD.51.2693}{\emph{Phys. Rev. D}
  {\bfseries 51} (1995) 2693}
  [\href{https://arxiv.org/abs/hep-ph/9410346}{{\ttfamily hep-ph/9410346}}].

\bibitem{Dolgov1997}
A.~Dolgov, K.~Freese, R.~Rangarajan and M.~Srednicki, \emph{Baryogenesis during
  reheating in natural inflation and comments on spontaneous baryogenesis},
  \href{https://doi.org/10.1103/PhysRevD.56.6155}{\emph{Phys. Rev. D}
  {\bfseries 56} (1997) 6155}
  [\href{https://arxiv.org/abs/hep-ph/9610405}{{\ttfamily hep-ph/9610405}}].

\bibitem{GarciaBellido1999}
J.~Garcia-Bellido, D.~Y. Grigoriev, A.~Kusenko and M.~E. Shaposhnikov,
  \emph{Nonequilibrium electroweak baryogenesis from preheating after
  inflation}, \href{https://doi.org/10.1103/PhysRevD.60.123504}{\emph{Phys.
  Rev. D} {\bfseries 60} (1999) 123504}
  [\href{https://arxiv.org/abs/hep-ph/9902449}{{\ttfamily hep-ph/9902449}}].

\bibitem{Davidson2000}
S.~Davidson, M.~Losada and A.~Riotto, \emph{A new perspective on baryogenesis},
  \href{https://doi.org/10.1103/PhysRevLett.84.4284}{\emph{Phys. Rev. Lett.}
  {\bfseries 84} (2000) 4284}
  [\href{https://arxiv.org/abs/hep-ph/0001301}{{\ttfamily hep-ph/0001301}}].

\bibitem{Megevand2001}
A.~Megevand, \emph{Effect of reheating on electroweak baryogenesis},
  \href{https://doi.org/10.1103/PhysRevD.64.027303}{\emph{Phys. Rev. D}
  {\bfseries 64} (2001) 027303}
  [\href{https://arxiv.org/abs/hep-ph/0011019}{{\ttfamily hep-ph/0011019}}].

\bibitem{Tranberg2003}
A.~Tranberg and J.~Smit, \emph{Baryon asymmetry from electroweak tachyonic
  preheating}, \href{https://doi.org/10.1088/1126-6708/2003/11/016}{\emph{JHEP}
  {\bfseries 11} (2003) 016}
  [\href{https://arxiv.org/abs/hep-ph/0310342}{{\ttfamily hep-ph/0310342}}].

\bibitem{Tranberg2006}
A.~Tranberg and J.~Smit, \emph{Simulations of cold electroweak baryogenesis:
  Dependence on Higgs mass and strength of CP-violation},
  \href{https://doi.org/10.1088/1126-6708/2006/08/012}{\emph{JHEP} {\bfseries
  08} (2006) 012} [\href{https://arxiv.org/abs/hep-ph/0604263}{{\ttfamily
  hep-ph/0604263}}].

\bibitem{Hertzberg2014}
M.~P. Hertzberg and J.~Karouby, \emph{Generating the observed baryon asymmetry
  from the inflaton field},
  \href{https://doi.org/10.1103/PhysRevD.89.063523}{\emph{Phys. Rev. D}
  {\bfseries 89} (2014) 063523}
  [\href{https://arxiv.org/abs/1309.0010}{{\ttfamily 1309.0010}}].

\bibitem{Hertzberg2014a}
M.~P. Hertzberg and J.~Karouby, \emph{Baryogenesis from the inflaton field},
  \href{https://doi.org/10.1016/j.physletb.2014.08.021}{\emph{Phys. Lett. B}
  {\bfseries 737} (2014) 34} [\href{https://arxiv.org/abs/1309.0007}{{\ttfamily
  1309.0007}}].

\bibitem{Hertzberg2014b}
M.~P. Hertzberg, J.~Karouby, W.~G. Spitzer, J.~C. Becerra and L.~Li,
  \emph{Theory of self-resonance after inflation. ii. quantum mechanics and
  particle-antiparticle asymmetry},
  \href{https://doi.org/10.1103/PhysRevD.90.123529}{\emph{Phys. Rev. D}
  {\bfseries 90} (2014) 123529}
  [\href{https://arxiv.org/abs/1408.1398}{{\ttfamily 1408.1398}}].

\bibitem{Lozanov2014}
K.~D. Lozanov and M.~A. Amin, \emph{End of inflation, oscillons, and
  matter-antimatter asymmetry},
  \href{https://doi.org/10.1103/PhysRevD.90.083528}{\emph{Phys. Rev. D}
  {\bfseries 90} (2014) 083528}
  [\href{https://arxiv.org/abs/1408.1811}{{\ttfamily 1408.1811}}].

\bibitem{Zurek2014}
K.~M. Zurek, \emph{Asymmetric dark matter: Theories, signatures, and
  constraints},
  \href{https://doi.org/10.1016/j.physrep.2013.12.001}{\emph{Phys. Rept.}
  {\bfseries 537} (2014) 91} [\href{https://arxiv.org/abs/1308.0338}{{\ttfamily
  1308.0338}}].

\bibitem{Xue2023}
S.-S. Xue, \emph{Massive particle pair production and oscillation in Friedman
  universe: its effect on inflation},
  \href{https://doi.org/10.1140/epjc/s10052-023-11195-6}{\emph{Eur. Phys. J. C}
  {\bfseries 83} (2023) 36} [\href{https://arxiv.org/abs/2112.09661}{{\ttfamily
  2112.09661}}].

\bibitem{Xue2023a}
S.-S. Xue, \emph{Massive particle pair production and oscillation in Friedman
  universe: reheating energy and entropy, and cold dark matter},
  \href{https://doi.org/10.1140/epjc/s10052-023-11524-9}{\emph{Eur. Phys. J. C}
  {\bfseries 83} (2023) 355}
  [\href{https://arxiv.org/abs/2006.15622}{{\ttfamily 2006.15622}}].

\bibitem{Xue2024}
S.-S. Xue, \emph{Holographic massive plasma state in Friedman universe:
  cosmological fine-tuning and coincidence problems}, {\emph{JCAP} 05(2024)113}
  [\href{https://arxiv.org/abs/2309.15488}{{\ttfamily 2309.15488}}].

\bibitem{Xue2015}
S.-S. Xue, \emph{How universe evolves with cosmological and gravitational
  constants},
  \href{https://doi.org/10.1016/j.nuclphysb.2015.05.022}{\emph{Nucl. Phys. B}
  {\bfseries 897} (2015) 326}
  [\href{https://arxiv.org/abs/1410.6152}{{\ttfamily 1410.6152}}].

\bibitem{PhysRevLett.67.2427}
Y.~Kluger, J.~M. Eisenberg, B.~Svetitsky, F.~Cooper and E.~Mottola, \emph{Pair
  production in a strong electric field},
  \href{https://doi.org/10.1103/PhysRevLett.67.2427}{\emph{Phys. Rev. Lett.}
  {\bfseries 67} (1991) 2427}.

\bibitem{Ruffini_2003}
R.~Ruffini, L.~Vitagliano and S.-S. Xue, \emph{On plasma oscillations in strong
  electric fields},
  \href{https://doi.org/10.1016/s0370-2693(03)00299-5}{\emph{Physics Letters B}
  {\bfseries 559} (2003) 12–19}.

\bibitem{Ruffini_2010}
R.~Ruffini, G.~Vereshchagin and S.-S. Xue, \emph{Electron–positron pairs in
  physics and astrophysics: From heavy nuclei to black holes},
  \href{https://doi.org/10.1016/j.physrep.2009.10.004}{\emph{Physics Reports}
  {\bfseries 487} (2010) 1–140}.

\bibitem{Wang2020}
Q.~Wang and W.~G. Unruh, \emph{Vacuum fluctuation, microcyclic universes, and
  the cosmological constant problem},
  \href{https://doi.org/10.1103/PhysRevD.102.023537}{\emph{Phys. Rev. D}
  {\bfseries 102} (2020) 023537}
  [\href{https://arxiv.org/abs/1904.08599}{{\ttfamily 1904.08599}}].

\bibitem{Wang2020a}
Q.~Wang, \emph{Reformulation of the cosmological constant problem},
  \href{https://doi.org/10.1103/PhysRevLett.125.051301}{\emph{Phys. Rev. Lett.}
  {\bfseries 125} (2020) 051301}
  [\href{https://arxiv.org/abs/1904.09566}{{\ttfamily 1904.09566}}].

\bibitem{Xue2020}
S.-S. Xue, \emph{Cosmological constant, matter, cosmic inflation and
  coincidence}, \href{https://doi.org/10.1142/S0217732320501230}{\emph{Mod.
  Phys. Lett. A} {\bfseries 35} (2020) 2050123}
  [\href{https://arxiv.org/abs/2004.10859}{{\ttfamily 2004.10859}}].

\bibitem{Xue2019}
S.-S. Xue, \emph{Cosmological $\Lambda$ driven inflation and produced massive
  particles},  \href{https://arxiv.org/abs/1910.03938}{{\ttfamily 1910.03938}}.

\bibitem{Kolb1990}
E.~W. Kolb and M.~S. Turner, \emph{The Early Universe}, vol.~69. Westview
  press, A member of the Perseus Books Group, 1990,
  \href{https://doi.org/10.1201/9780429492860}{10.1201/9780429492860}.

\bibitem{Lee1977}
B.~W. Lee and S.~Weinberg, \emph{Cosmological lower bound on heavy neutrino
  masses}, \href{https://doi.org/10.1103/PhysRevLett.39.165}{\emph{Phys. Rev.
  Lett.} {\bfseries 39} (1977) 165}.

\bibitem{1999A&A...350..334R}
R.~{Ruffini}, J.~D. {Salmonson}, J.~R. {Wilson} and S.~S. {Xue}, \emph{{On the
  pair electromagnetic pulse of a black hole with electromagnetic structure}},
  \href{https://doi.org/10.48550/arXiv.astro-ph/9907030}{\emph{Astron.
  Astrophys} {\bfseries 350} (1999) 334}
  [\href{https://arxiv.org/abs/astro-ph/9907030}{{\ttfamily
  astro-ph/9907030}}].

\bibitem{Ruffini1999b}
R.~Ruffini, J.~D. Salmonson, J.~R. Wilson and S.~S. Xue, \emph{On the evolution of
  the pair-electromagnetic pulse of a charged black hole},
  \href{https://doi.org/10.1051/aas:1999330}{\emph{Astron. Astrophys 138,
  511-512 (1999)} {\bfseries 138} (1999) 511}
  [\href{https://arxiv.org/abs/astro-ph/9905021}{{\ttfamily
  astro-ph/9905021}}].

\bibitem{2000A&A...359..855R}
R.~{Ruffini}, J.~D. {Salmonson}, J.~R. {Wilson} and S.~S. {Xue}, \emph{{On the
  pair-electromagnetic pulse from an electromagnetic black hole surrounded by a
  baryonic remnant}},
  \href{https://doi.org/10.48550/arXiv.astro-ph/0004257}{\emph{Astron.
  Astrophy} {\bfseries 359} (2000) 855}
  [\href{https://arxiv.org/abs/astro-ph/0004257}{{\ttfamily
  astro-ph/0004257}}].

\bibitem{Sakurai2020}
J.~J. Sakurai and J.~Napolitano, \emph{Modern quantum mechanics}. Cambridge:
  Cambridge University Press, 3rd revised edition~ed., 2020,
  \href{https://doi.org/10.1017/9781108587280}{10.1017/9781108587280}.

\bibitem{landaunon}
L.~Landau and E.~Lifshiz, \emph{Quantum Mechanics (Non-relativistic Theory)}.
  Elsevier Butterworth-Heinemann, 1977.

\bibitem{Hollands2002}
S.~Hollands and R.~M. Wald, \emph{An alternative to inflation},
  \href{https://doi.org/10.1023/A:1021175216055}{\emph{Gen. Rel. Grav.}
  {\bfseries 34} (2002) 2043}
  [\href{https://arxiv.org/abs/gr-qc/0205058}{{\ttfamily gr-qc/0205058}}].

\bibitem{Mielczarek2011}
J.~Mielczarek, \emph{Reheating temperature from the CMB},
  \href{https://doi.org/10.1103/PhysRevD.83.023502}{\emph{Phys. Rev. D}
  {\bfseries 83} (2011) 023502}
  [\href{https://arxiv.org/abs/1009.2359}{{\ttfamily 1009.2359}}].

\end{thebibliography}


\providecommand{\href}[2]{#2}\begingroup\raggedright\endgroup

\end{document}